\documentclass[aps,prb,twocolumn,showpacs,preprintnumbers,letterpaper]{revtex4}

\usepackage{multirow}
\usepackage{setspace}
\usepackage{braket}
\usepackage{amsmath}
\usepackage[pdftex]{graphicx}
\usepackage{color}

\begin{document}
\DeclareGraphicsExtensions{.pdf,.png,.gif,.jpg,.eps}
\title{Double Quantum Dots in Carbon Nanotubes}
\author{J. von Stecher,$^*$ B. Wunsch,$^\dag$ M. Lukin,$^{\dag}$  E. Demler,$^{\dag}$ and A. M. Rey$^{*}$}
\affiliation{$^*$JILA, University of Colorado and National Institute of Standards and Technology,
Boulder, Colorado 80309-0440,}
\affiliation{$^\dag$Physics Department, Harvard University, Cambridge, Massachusetts, 20138,}
\date{\today}

\begin{abstract}
We study the two-electron
 eigenspectrum of a carbon-nanotube double quantum dot with spin-orbit coupling. Exact calculation are combined with a simple model to provide an intuitive and accurate description of single-particle and interaction effects.
For symmetric dots and weak magnetic fields, the two-electron ground state is antisymmetric in the spin-valley degree of freedom and is not  a pure spin-singlet state. When double occupation of one dot is favored by increasing the detuning between the dots, the Coulomb interaction causes strong correlation effects realized by higher orbital-level mixing.
Changes in the double-dot configuration affect the relative strength of the electron-electron interactions and can lead to different ground state transitions. In particular, they can favor a ferromagnetic ground state both in spin and valley degrees of freedom.
 The strong suppression of the energy gap  can cause the disappearance of the Pauli blockade in transport experiments and thereby can also limit
the stability of spin-qubits in quantum information proposals.
Our analysis is generalized to an array of coupled dots which is expected to exhibit rich many-body behavior.

\end{abstract}

\maketitle

\section{Introduction}
Experiments on few-electron double quantum dots allow the measurement and manipulation of the spin degree of freedom of the confined electrons\cite{Hanson2007}.  Such control  is at the heart of semiconductor-based spintronics~\cite{wolf2001spintronics,awschalom2007challenges} and quantum-information proposals~\cite{loss1998quantum,cerletti2005recipes}.
Recently, substantial experimental efforts have been focussed on controlling electrons in carbon nanotube double quantum
dots, and many of the capabilities previously achieved in GaAs double dots\cite{Ono2002,petta2005cmc,Hanson2007,Koppens06}
are starting to be reproduced\cite{Churchill09,Churchill09b,Gotz09}. These include the
ability to start from an empty double dot and systematically fill it with electrons.
 Since $^{12}$C has no nuclear spin, carbon based nanostructures are expected to
   reduce hyperfine induced decoherence as compared with GaAs\cite{bulaev2008soi}.

Furthermore, carbon based materials exhibit richer physics than GaAs semiconducting materials because of the additional valley degree of freedom\cite{jarillo2005orbital,oreg2000spin,moriyama2005four}. In principle, the spin and valley degrees of freedom could lead to a SU(4) symmetry at zero magnetic field  instead of the standard SU(2) symmetry in conventional semiconductors (see e.g. Ref.~\onlinecite{egger2009kondo} and references therein). However, it has been recently demonstrated~\cite{kuemmeth2008csa}
that the enhancement of the spin-orbit splitting in small-radius nanotubes breaks the four-fold degeneracy of the single-electron ground state into a two-fold degeneracy.

In this work, we study how spin-orbit coupling and electron-electron interaction effects are manifested in the two-electron spectrum and transport properties of a carbon nanotube double dot. This represents an extension of previous studies on few-electron physics in a single carbon nanotube dot\cite{wunsch09,secchi2009B,secchi2009cvs}.  We find that, despite of spin-orbit coupling and the existence of an additional valley degree of freedom, the two-electron  eigenstates  can be 
 separated in an orbital part and a spin-valley part that are, to a very good approximation, independent of each other. The spin-valley part can be grouped in  six antisymmetric and  ten symmetric spin-valley eigenstates which we refer to as multiplets.

The separation of spin-valley degrees of freedom significantly simplifies the description of the systems and allows us to draw analogies with standard GaAs double dots. Our main results can be summarized as follows: (a) For
 dots at zero magnetic field and no detuning, each dot is populated by a single electron and tunneling is suppressed because of Coulomb interactions. Thus, interdot coupling only occurs virtually via superexchange interactions that determine the ground state symmetry. In this regime, we find  a spin-valley antisymmetric ground  state,
that does not have a well defined-spin due to the spin-orbit coupling.
(b) For large detuning between the dots, double occupation of the same dot becomes favorable. In this regime, Coulomb interactions  can mix higher orbitals in the two-electron ground state\cite{wunsch09,secchi2009B,secchi2009cvs}. This admixture significantly reduces the energy spacing
between the multiplets  (which, for weakly interacting electrons, is determined by the orbital level spacing).
The interplay between spin-orbit coupling and interaction then leads to a ferromagnetic ground state above a small critical magnetic field, since the Zeeman terms overcome the strongly suppressed splitting between effective singlet an triplet. c) The reduction  of the energy gap between orbitally symmetric and antisymmetric states caused by the Coulomb interaction affects transport properties through the dot and can lead to the disappearance of the so-called  Pauli-blockade (suppression of current through the double dot due to the Pauli exclusion principle)
and might explain the absence of Pauli blockade reported in Ref.~\onlinecite{Gotz09}.  The absence of spin-blockade might affect the performance of quantum information proposal which use spin-qubits in double dots, since gate operation in those proposals is based on the spin-blockade mechanism. The disappearance of Pauli blockade might be prevented by reducing Coulomb correlations by either working with short dots, or by covering the nanotube by large dielectrics\cite{wunsch09}.


This paper is organized as follows. In the section II, we introduce the microscopic model for the double dot and analyze the non-interacting predictions taking into account a  magnetic field parallel to the nanotube axis,  spin-orbit couplings and detuning between the dots. Then, we construct a simple two-electron model that captures the interaction effects.  This model is then compared with solutions of an exact many-band Hamiltonian using localized single-particle orbitals.
In Section~\ref{results}, we discuss the energy spectrum of two interacting electrons in a double dot in three different detuning regimes corresponding to (i) a symmetric double dot, with one electron in each dot, (ii) strong detuning, with both electrons in the same dot, and (iii) at the crossover between both regimes. We then analyze the transport properties of the double dot.
Finally, we discuss how to extend our low energy behavior analysis to serially coupled quantum dots.
In section IV, we present the conclusion. Technical details on the calculation of Coulomb matrix elements and the derivation of the rate equations used for transport are presented in Appendices~\ref{App:CoulMat} and~\ref{rate}.
\begin{figure}[h]
\begin{center}
\begin{tabular}{c}
\includegraphics[scale=0.23,angle=0]{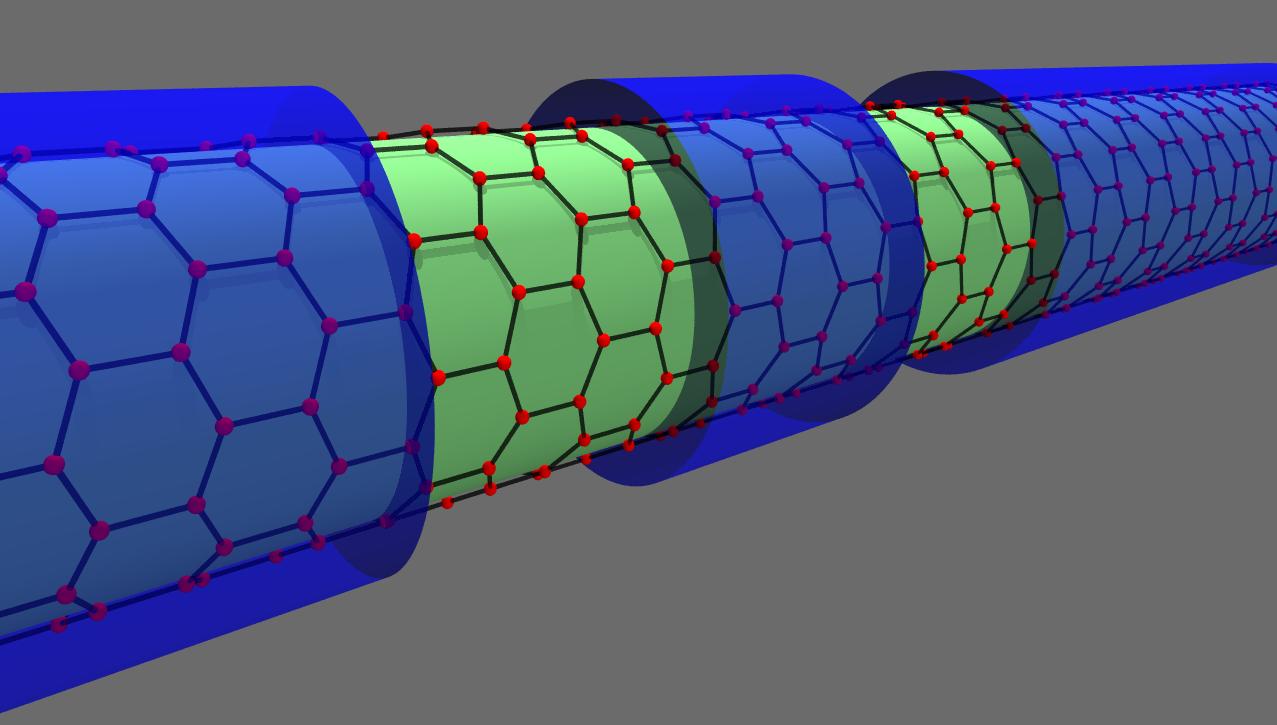}\\
\includegraphics[scale=0.3,angle=0]{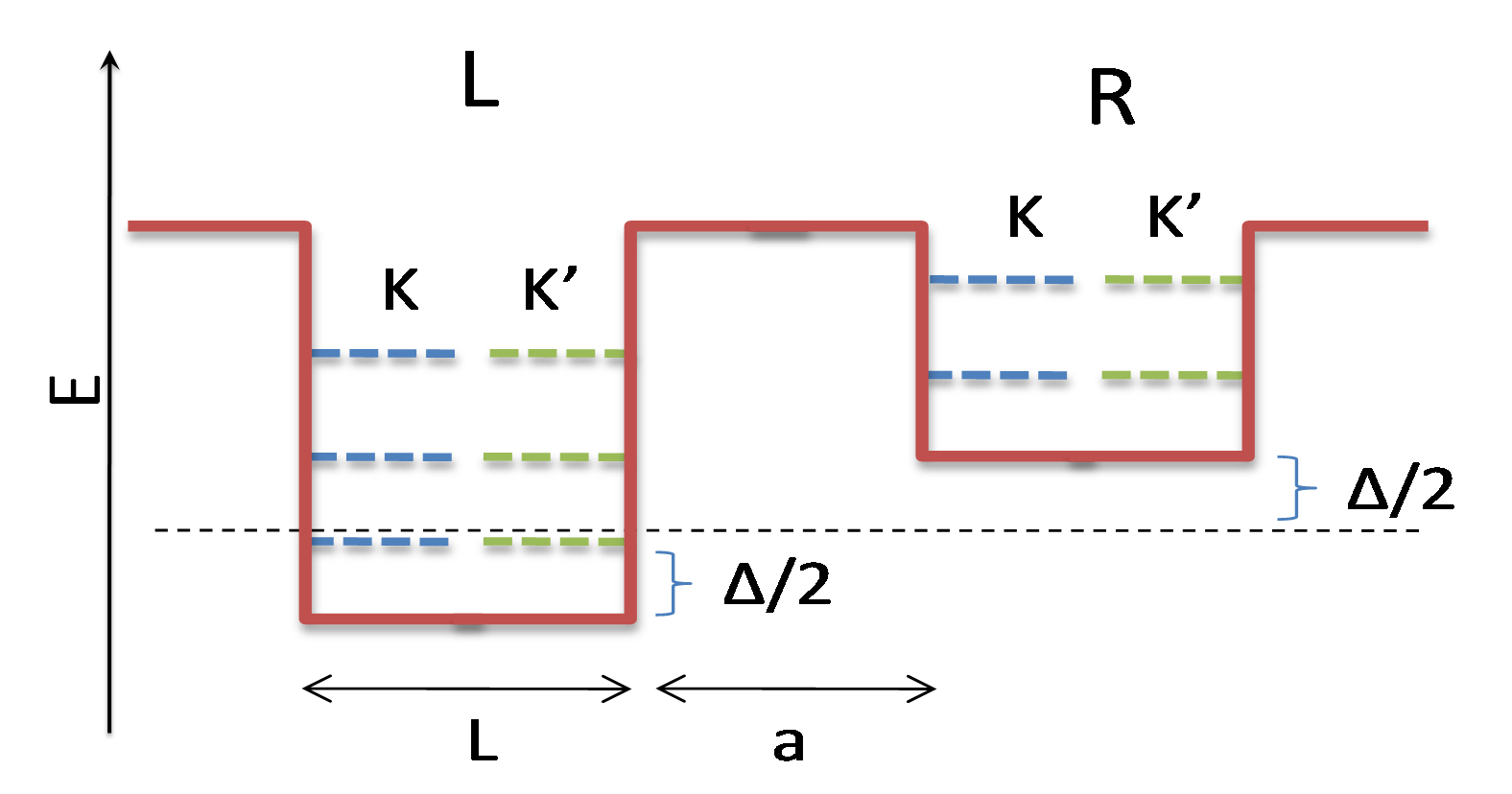}
\end{tabular}
\caption{(Color online) Schematic representations of a double quantum dot in carbon nanotube. Top: Blue regions correspond to the potential barriers. Bottom: Double dot potential at finite detuning $\Delta$. Dashed lines schematically represent the single-particle orbitals.}
\label{dots3d}
\end{center}
\end{figure}

\section{Model}
Single-wall carbon nanotubes are formed by a single layer of graphite called graphene rolled up into a cylinder.
Graphene has a honeycomb lattice formed by covalently bonded carbon atoms. Its electronic properties  are determined by the $p_z$ orbital of the carbon atom.
The low energy spectrum of graphene consists of two Dirac cones located at the $K$ and $K'=-K$ points of the graphene's Brillouin zone, where the valence and conduction bands touch. To characterize the two Dirac cones, we introduce the valley index $\tau=\pm1$, where $\tau=1$ corresponds to the $K'$ point and $\tau=-1$ to the $K$ point. The behavior of graphene in the presence of an external potential can be described by an effective mass approximation, or $\vec{\kappa}\cdot\vec{p}$ theory~\cite{divincenzo1984sce}. In this approximation, the envelope wave function for the $A$ and $B$ sites of the two-atom unit cell in a honeycomb lattice follows an effective Dirac equation.
In a carbon nanotube, the cylindrical structure imposes a quantization condition that leads to either metallic or semiconducting nanotubes, depending on the orientation of the underlying lattice with respect to the symmetry axis of the tube. Here, we will focus on the behavior of semiconducting nanotubes.

\subsection{Single particle spectrum and interactions}\label{SParticle}
In this section we follow previous work\cite{bulaev2008soi,trauzettel2007sqg,wunsch09} to derive the localized eigenstates of a semiconducting nanotube with an additional confinement potential along the tube, which is controlled by external gates.
We describe the confinement  potential of each dot by a square well\cite{bulaev2008soi,trauzettel2007sqg} (see Fig.~\ref{dots3d}). The form of the confinement potential will not affect our results qualitatively and we note that for a single dot, the results for parabolic confinement and square well are in good agreement\cite{secchi2009cvs,wunsch09}.  The external potential leads to a discrete set of bound states.  Taking $\zeta$ as the direction of the nanotube axis and $\varphi$  as the angle perpendicular to the nanotube axis, we can write the
single particle Hamiltonian as
\begin{equation}
  \label{H01d}
H_0=-i \hbar v (\tau\sigma_1\frac{1}{R}\partial_\varphi+ \sigma_2\partial_\zeta)+V(\zeta),
\end{equation}
where $v$ is the Fermi velocity, $\sigma_i$ are the Pauli matrices operating over the sublattice space, and $V(\zeta)$ is the external potential that describes one or two dots.
 The eigenstates are determined by matching the solutions for the dot and barrier regions, which are of the form:
\begin{equation}
  \label{Psi}
\Psi^{\tau,\kappa,k}(\varphi,\zeta)=e^{i(\kappa R\varphi+k\zeta)}\left(\begin{array}{c} z^{\tau}_{\kappa,k}\\1 \end{array}\right).
\end{equation}
Here $\kappa,k$ denote the wave vectors around and along the tube, $z^{\tau}_{\kappa,k}=\pm(\tau\kappa-ik)/\sqrt{\kappa^2+k^2}$, and the energy is given by $E_{\kappa,k}=\pm\hbar v\sqrt{\kappa^2+k^2}$. Solving the effective Dirac equation for $V_{1D}(\zeta)$, which  is 0 for $0<\zeta<L$ and $V_g$ otherwise,  leads to a
quantization condition for the longitudinal momentum modes $k_n$ of the localized states, where $n$ denotes the band index\cite{bulaev2008soi}.

So far, the $K$ and $K'$ solutions are degenerate and independent of spin, $\sigma=\uparrow\downarrow$, leading to a four-fold symmetry. However, this symmetry is broken by spin-orbit coupling corrections\cite{kuemmeth2008csa,ando2000soi,Paco06}  and by a constant magnetic field $B$ along the nanotube axis $\widehat{\zeta}$.
The spin-orbit coupling and the presence of a magnetic field modify the quantization condition in the $\widehat{\varphi}$ direction,
\begin{eqnarray}
\kappa&=&\kappa_0+\Phi_{AB}/\Phi_0 R-s\Delta_{SO}/(\hbar v)\\
\kappa_0&=&\tau/3R.\label{eq:kappa}
\end{eqnarray}
 Here, $s=\pm1/2$ is the quantum number corresponding
to the spin operator parallel to the nanotube ($\hat{S}_\zeta\ket{\uparrow}=1/2\ket{\uparrow}$ and $\hat{S}_\zeta\ket{\downarrow}=-1/2\ket{\downarrow}$); $\Delta_{SO}\approx1\mbox{meV}/R$[nm] is the energy splitting due to spin-orbit coupling, $\Phi_{AB}=B \pi R^2$ is the Aharonov-Bohm flux through the nanotube\cite{minot2004determination} and $\Phi_0=hc/|e|$. The magnetic field also leads to a spin Zeeman term $H_z=s\hbar\omega$, where $\omega=|e| g B/(2m_0 c)$ is the Zeeman frequency in terms of the gyromagnetic constant $g$, the electron mass $m_0$, and the speed of light $c$.
The sign convention for the spin used here is opposite to the one of Ref.~\onlinecite{wunsch09}.
 Note that in Eq.~\eqref{eq:kappa} we only consider the lowest mode in the transverse direction, since excitations involve energies of about $\hbar v/R$ which are much larger than longitudinal-excitation energies or Coulomb-interaction effects as long as $R \ll L$.

\begin{figure}[h]
\begin{center}
\includegraphics[scale=0.7,angle=0]{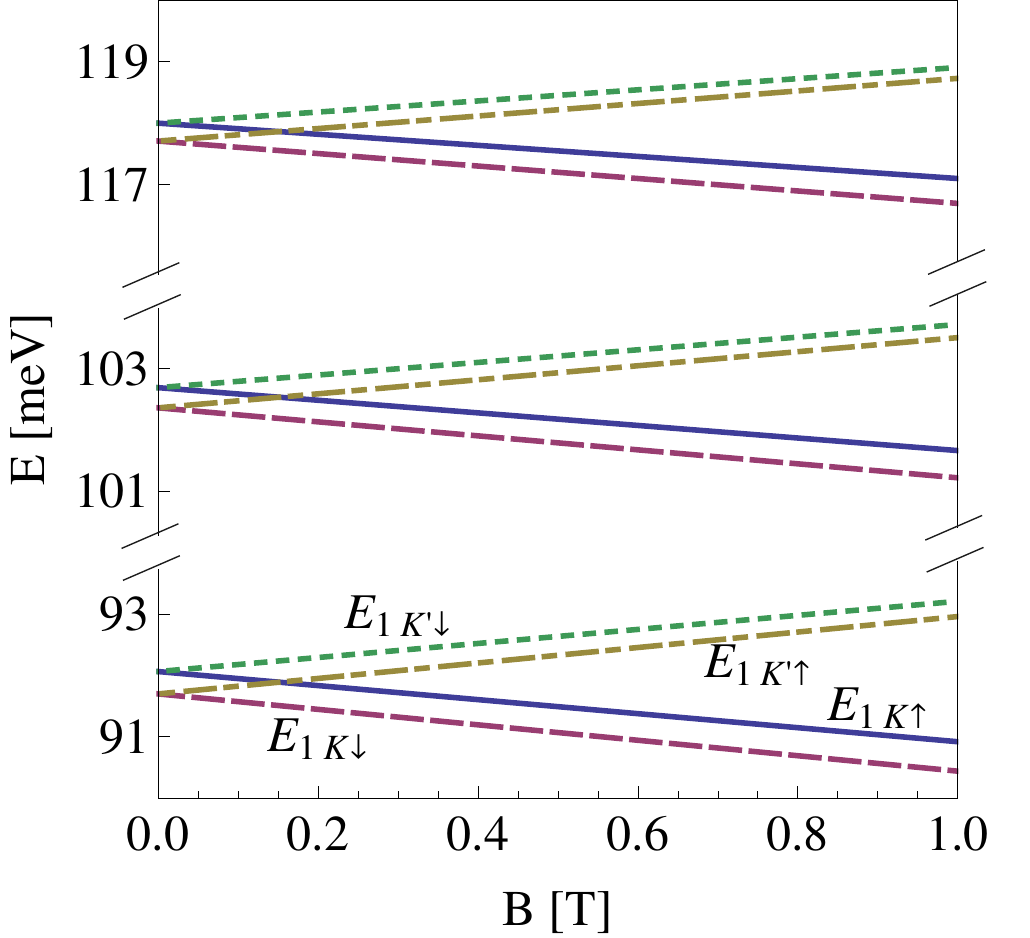}
\caption{(Color online) Single particle spectrum as a function of $B$ for $L=70$~nm, $V_g=78$~meV, and $R=2.5$~nm and $\Delta_{SO}<0$. Solid curves correspond to $E_{nK\uparrow}$, dashed  curves correspond to $E_{nK\downarrow}$, dash-dotted curves  correspond to $E_{nK'\uparrow}$, and dotted curves correspond to $E_{nK'\downarrow}$, where $n=1,2, 3$ is the band label.}
\label{E1multi}
\end{center}
\end{figure}

An example of the single particle energy spectrum of a single dot is shown in Fig.~\ref{E1multi}. Note that we measure the energy with respect to the center of the gap, so that the dominant part of the single-particle energy is constant and given by $\hbar v \kappa\approx 220 \,\mbox{meV}/R$[nm].
Generally, disorder or the confinement potential itself can lead to intervalley coupling. However, for a noticeable effect, the potential must change on the scale of the nearest neighbor lattice spacing between carbon atoms $a_0=1.4$~\AA. The experiment of Ref.~\onlinecite{kuemmeth2008csa} shows only a tiny valley mixing, and we neglect intervalley scattering in this work.

The longitudinal wave vector depends indirectly on spin-valley quantum numbers and on magnetic field and these effects are included in the multi band calculations.
However, since $\kappa_0\gg\Phi_{AB}/(\Phi_0 R),\Delta_{SO}/(2\hbar v),k_n$, the single particle energies can be significantly simplified, yielding
\begin{eqnarray}
E_{n,\alpha}&\approx&  E^c_n(V_g)+E_{\alpha}\label{E1pA}\\
E_{\alpha}&=&B(\tau\mu_{orb}+2 \mu_{spin} s)-\Delta_{SO} \tau s\,. \label{Etaus}
\end{eqnarray}
Here we have introduced a single quantum number $\alpha\equiv (\tau,s)$ to describe the spin valley degrees of freedom ($\alpha=K\uparrow ,K\uparrow ,K'\downarrow ,K'\downarrow $). The confinement energy $E^c_n(V_g)=\hbar v\sqrt{\kappa_0^2+k_n^2}$ is the single-particle spectrum of a dot with potential depth $V_g$ ignoring magnetic field, spin, and spin-orbit dependences;  $\mu_{spin}=\hbar\omega/(2B)=\hbar |e| g /(4m_0 c)$; and
$\mu_{orb}=( \pi R|e|/hc)$.
Within this approximation the orbital part of the single-particle states separate from the spin-valley part.
Figure~\ref{E1multi} illustrates that, for the parameters studied in this work, spin-valley splitting is basically the same for all longitudinal bands, and Eq.~\eqref{E1pA} provides a good description of the single-particle spectrum of a single dot.

Next, we consider a biased double-dot system, schematically presented in Fig.~\ref{dots3d}. In experiment the double quantum dot is formed by applying appropriate voltages to external gates and we model the resulting confinement potential $V_{2D}(\zeta)$ by a square well potential that is $-\Delta/2$ for $-a/2-L<\zeta<-a/2$, $\Delta/2$ for $a/2<\zeta<a/2+L$, and $V_g$ otherwise. The length of the dots is $L$ and $a$ is the width of the interdot barrier. As discussed previously we do not expect our results to change qualitatively, if a smoother potential is used.
In the double-dot system at finite detuning, the depths of the dots change affecting the single particle energies.
In the numerical calculation, we determine the eigenspectrum of $V_{2D}$ exactly.
The main effect of the detuning is an energy shift $\pm \Delta/2$ to the single-particle eigenenergies, where ``$+$'' corresponds to right dot, and ``$-$'' to the left dot.
Using  Eq.~\eqref{E1pA} and neglecting interdot tunneling, the energies of the localized left and right single-particle orbitals are approximately
\begin{equation}
  \label{E1pA2}
E^{R/L}_{n,\alpha}\approx E^c_n(V_g \pm\Delta/2)+E_{\alpha}\approx E^c_n(V_g)\pm \Delta/2+E_{\alpha}.
\end{equation}

When more than one electron is confined in the single or double dot, electron-electron interactions become important.
The electrons interact through the long-range Coulomb potential
\begin{equation}
V_c({\bf r_1},{\bf r_2})=\frac{e^2}{k_d|{\bf r_1}-{\bf r_2|}},
\label{coulomb}
\end{equation}
 where $k_d$ denotes the dielectric constant.
Coulomb interactions allow for certain off-diagonal matrix elements in valley space that are produced by intervalley scattering\cite{egger1997ele}.
 However, these matrix elements are small for quantum dots with a size much larger than the interatomic distance; they are neglected in this work\cite{wunsch09}.

To obtain an accurate description of interacting few electron systems, we extend the single-dot treatment of Refs.~\onlinecite{wunsch09,secchi2009B} to the double dot system. We construct single particle orbitals localized in the left and right dots from the exact single particle solutions of the double dot and then we use
 them to expand the many-body Hamiltonian (see more details in Appendix~\ref{App:CoulMat}). The single particle orbitals have a weak dependence on the spin-valley degrees of freedom that comes from the dependence of the wave vectors $\kappa$ and $k$ on $\tau$ and $s$. This leads to a dependence of the interaction matrix elements on the spin-valley degrees of freedom. However, this dependence is very weak and, to a very good approximation, can be neglected. Thus, interactions can be considered spin-valley independent allowing the separation of the orbital and the spin-valley contributions in the two-electron solutions~\cite{wunsch09}.

\subsection{Separating orbital from spin-valley degrees of freedom}

\begin{figure}[h]
\begin{center}
\begin{tabular}{cc}
\includegraphics[scale=0.85,angle=0]{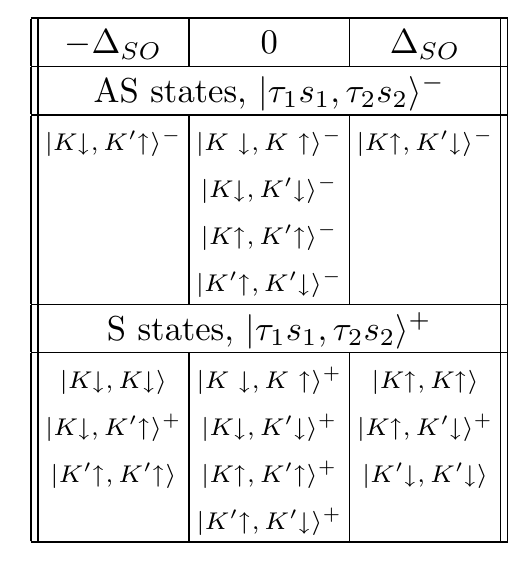} &\includegraphics[scale=0.38,angle=0]{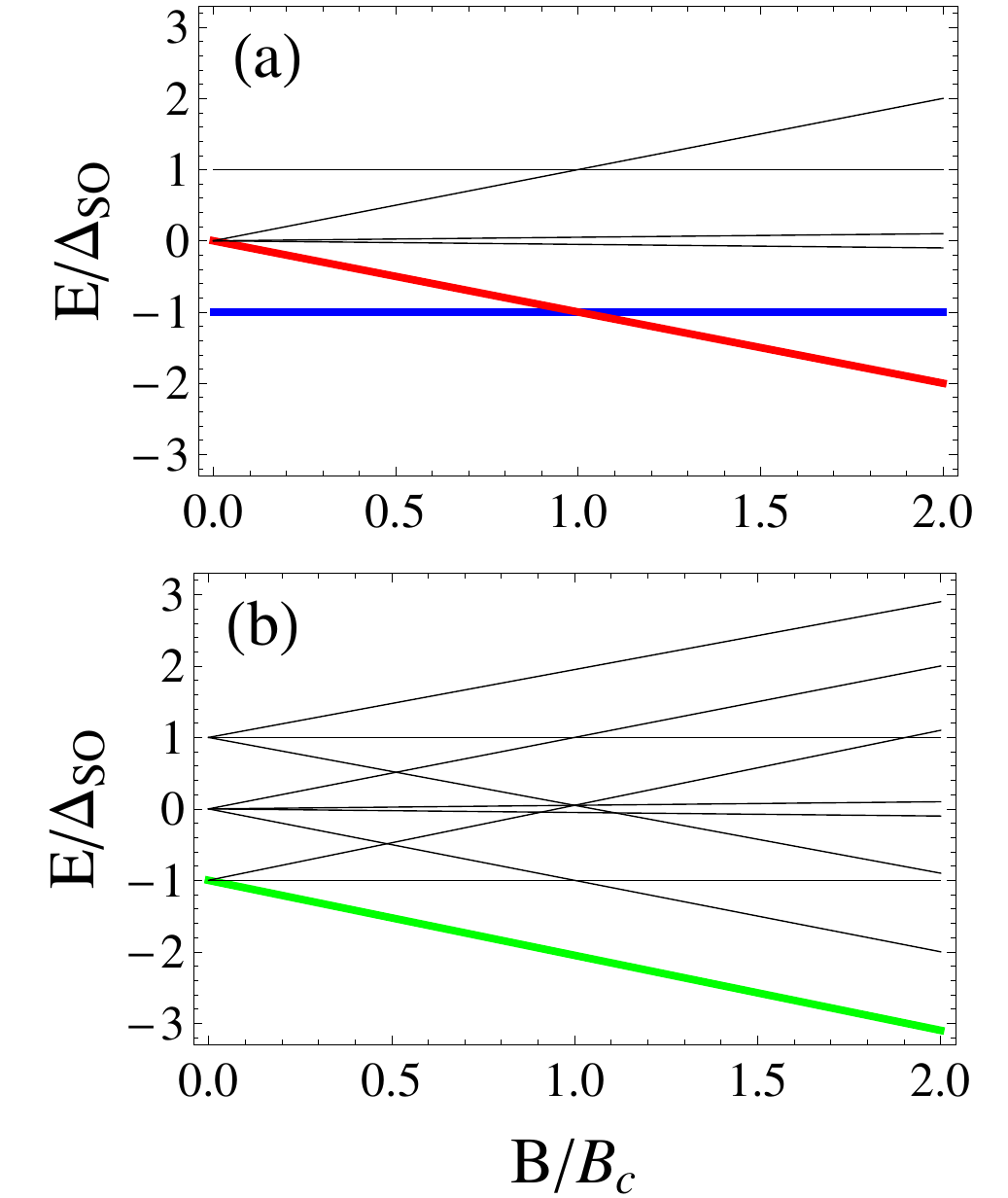}
\end{tabular}
\caption{Table: spin-valley multiplets, here {\scriptsize $ |\tau_1 s_1,\tau_2s_2\rangle^{\pm}= (|\tau_1s_1,\tau_2s_2\rangle \pm  |\tau_2s_2,\tau_1s_1\rangle)/\sqrt{2}$}. The states in each multiplet are grouped in three columns according to their spin-orbit energy. Figures: Schematic magnetic-field dependence of  antisymmetric (a) and symmetric (b) multiplets.}
\label{table1}
\end{center}
\end{figure}

Since interactions can be considered diagonal in spin-valley degrees of freedom, orbital and spin-valley part of the two-electron solutions provide independent contributions to the energies and the wave functions~\cite{wunsch09}.
The total two-particle wave function must be  antisymmetric with respect to particle exchange and the symmetry of the orbital part must always be opposite to that of the spin-valley part. Thus, the two-particle spectrum can be grouped according to their orbital symmetry or parity under particle exchange in multiplets of six states if the orbital part is symmetric or ten states if the orbital part is antisymmetric.
The energy splitting between different multiplets, called $\epsilon$, depends on the orbital part, which is determined by electron-electron interactions and longitudinal confinement, and it is generally given by a correlated state that is represented as a superposition of various two-electron orbital wave functions.
The energy relations within a multiplet are exclusively determined by $E_\alpha$ that includes the orbital and spin Zeeman terms  as well as the spin-orbit coupling [Eq.~\eqref{Etaus}]. Therefore, the spin-valley part of the wave function always has the simple form shown in Table~\ref{table1}.

The magnetic field dependence of the two multiplets is illustrated in Fig.~\ref{table1}.
The competition between spin-orbit coupling and orbital Zeeman energy leads to a ground state crossings in the multiplet with antisymmetric spin-valley part at a critical magnetic field $B_c=\Delta_{SO}/2\mu_{orb}$. For $B<B_c$ the ground state of the antisymmetric spin-valley multiplet [Fig.~\ref{table1}~(a)] is in a superposition of spin singlet and triplet, since $\ket{K\downarrow, K'\uparrow}^-=(\ket{K K'}^+ \ket{\downarrow\uparrow}^-+\ket{K K'}^- \ket{\downarrow\uparrow}^+)/\sqrt{2}$. For $B<B_c$, the lowest state of the antisymmetric spin-valley multiplet $\ket{K\uparrow, K\downarrow}^-$ is antiferromagnetic in spin but ferromagnetic (or polarized) in valley space. The lowest state of the symmetric spin-valley multiplet $\ket{K\downarrow, K\downarrow}^+$ is ferromagnetic in both spin and valley space for all positive magnetic fields.

The energy difference between the different multiplets $\epsilon$ and the magnetic field $B$ determines the spin-valley symmetry of the ground state. The thick curves in Figs.~\ref{table1}(a) and (b) correspond to the possible ground states which we label according to their spin and valley symmetry.

\subsection{Model Description}

One of the objectives of this study is to be able to describe the evolution of the spectrum as the detuning is changed from small to large and the low energy configurations change from $|1L,1R\rangle$, i.e. one electron per dot, to $|2L,0R\rangle$ or $|0L,2R\rangle$, two electrons in the same dot. An accurate description of the strong interaction effects requires the inclusion of several single particle bands of the double dot system. For example, the behavior of the $|2L,0R\rangle$ configurations is expected to be very similar to that of the doubly occupied single dot and in that situation the strong correlations need to be described by many single particle orbitals~\cite{wunsch09,secchi2009B}. This situation makes the description and the interpretation of the results not very intuitive. However, we can significantly simplify the description by realizing that no matter how strongly correlated the system is, the parity is a good quantum number and the states can be classified according to the state parity. Thus, we can model the exact system with a simple effective Hamiltonian in the charge degrees of freedom that captures this dependence on parity and describes the energetically lowest multiplets of the $|1L,1R\rangle$, $|2L,0R\rangle$ and $|0L,2R\rangle$ configurations.

The charge degrees of freedom of two-electrons in a double dot can have three configurations: $|2L,0R\rangle^\pm$, $|1L,1R\rangle^\pm$,  and $|0L,2R\rangle^\pm$,  where $\pm$ characterizes the conserved orbital symmetry (i.e. + for antisymmetric spin-valley states and - for symmetric spin-valley states).
 Within this model, interaction effects can be obtained by diagonalizing the effective Hamiltonians $H_S$ and $H_{AS}$, where $S$ and $AS$ denote symmetric and antisymmetric spin-valley state.
The complex multiband problem is then reduced to simple 3 times three matrices.
\begin{eqnarray}
H_S=\left(
\begin{array}{ccc}
V+V_{ex}-\Delta & -t_{S}&0\\
-t_{S}&V_{LR} & t_{S}\\
0 &t_{S}&V+V_{ex}+\Delta
\end{array}
\right)
\label{HmodelS}
\end{eqnarray}
and
\begin{eqnarray}
H_{AS}=\left(
\begin{array}{ccc}
V-V_{ex}-\Delta & -t_{AS}&0\\
-t_{AS}&V_{LR} & -t_{AS}\\
0 &-t_{AS}&V-V_{ex}+\Delta
\end{array}
\label{HmodelAS}
\right).
\end{eqnarray}
These effective Hamiltonians include the onsite and nearest neighbor interactions $V$ and $V_{LR}$, the tunnelings in symmetric $t_S$ and antisymmetric $t_{AS}$ configuration and the detuning effects $\Delta$. The dependence of the interaction on the symmetry is introduced by an effective exchange term $V_{ex}$ that favors the antisymmetric spin-valley configuration. Equations~\ref{HmodelS} and \ref{HmodelAS} describe the energy related with the orbital part of the wavefunction.  The total energy also contains the contribution of the spin-valley part $E_{SV}=\sum_{\alpha}[E^c_1(V_g)+E_{\alpha}] n_{\alpha}$, where $E_{\alpha}$ was defined in Eq.\eqref{Etaus} and $n_{\alpha}$ denotes the occupation of states with spin-valley $\alpha$.

To gain qualitative understanding of the interaction and tunneling terms in the effective Hamiltonian, we analyze the limiting behaviors of the low-energy spectrum. First, we consider the limit of zero detuning and large local Coulomb interactions and we obtain that the lowest symmetric and antisymmetric spin-valley multiplets have energies:
 \begin{eqnarray}
E^{g}_{AS}&\approx&E_{SV}+V_{LR}-\frac{2t_{AS}^2}{V-V_{LR}-V_{ex}} +...,
\label{E11ap1}\\
E^{g}_S&\approx&E_{SV}+V_{LR}-\frac{2t_{S}^2}{V-V_{LR}+V_{ex}} +....
\label{E11ap2}
\end{eqnarray}
In the strong interaction regime, our numerical calculations indicate that, to a good approximation $t_S\approx t_{AS}\approx t_{12}$. This approximation allows to obtain  a simple expression for the energy splitting between different multiplets, $\epsilon=E^{g}_S-E^{g}_A\approx 4 t_{12}^2 V_{ex}/(V-V_{LR})^2$ in the limit of zero detuning and for $V_{ex}\ll V-V_{LR}$.

For a biased double dot system, the single particle energies acquire an energy shift of $\pm\Delta/2$ and in the limit of large detuning the two electrons occupy the same dot. In this limit, energies of the lowest symmetric and antisymmetric spin-valley multiplets are
\begin{eqnarray}
E^{g}_{AS}&\approx&E_{SV}-\Delta+V-V_{ex}+...,
\label{E20ap1}\\
E^{g}_S&\approx&E_{SV}-\Delta+V+V_{ex}+....
\label{E20ap2}
\end{eqnarray}
Thus, the energy splitting $\epsilon=E^{g}_S-E^{g}_A\approx 2 V_{ex}$ in the (2,0) configuration is mainly controlled by the exchange mechanism.

This effective model allows a simple and intuitive understanding of the underlying physical behavior of the double dot system. However, to extract the parameters $V$, $V_{LR}$, $V_{ex}$ and $t_{12}$ we need to solve exactly the single and double dot system.
In the next Section, we analyze the behavior of the double-dot system at different magnetic field, detuning and interaction regimes by comparing the exact diagonalization solutions discussed in the previous subsection with the model Hamiltonian. From this comparison, we  extract the parameters of the model.

\section{Results}\label{results}

In this section, we analyze the energy spectrum of two interacting electrons in a double dot in three different detuning regimes: (A) small detuning, with one electron in each dot; (B) strong detuning, with both electrons in the same dot and (C) at the crossover between both regimes. In subsection~\ref{subsecST}, we study the transport properties of the double dot.

An important energy scale of the two-particle spectrum is the energy spacing $\epsilon$ between the lowest energy states with an antisymmetric orbital part and the lowest state with a symmetric orbital part at zero magnetic field $B=0$.
 In all detuning regimes, we find $\epsilon>0$, in agreement  with the Lieb Mattis theorem\cite{Lieb61}. However, Coulomb correlations can significantly reduce $\epsilon$.

In our analysis, we have considered different double dot configurations by changing the length and depth of the dots as well as the interdot distance and found the same scaling of interaction effects as discussed in Ref.\onlinecite{wunsch09}. In this section we present results for the parameters $L=70$~nm, $a=20$~nm, $V_g=78$~meV, $R=2.5$~nm.  A single well with these parameters supports five bound states and has an energy splitting between the lowest two of them of $\hbar \omega_0\approx10.6$~meV. The dielectric constant $k_d$ is varied between ($1.5\le k_d\le3.5$) which allows us to explore the strongly interacting regime where new transitions  occur. Experimentally, however, it is easier to change the length of the dots. The parameter that characterize the strength of the interactions is the ratio $U/(\hbar\omega_0)$ where $U$ is the characteristic intradot interaction energy $U=e^2/k_d L$. This ratio is typically  between $1<U/(\hbar\omega_0)<5$  implying  a moderate/strong interaction regime. Another relevant parameter that determines the spin-valley nature of the ground state is, as we will discuss below, the ratio $2V_{ex}/\Delta_{SO}$ that reflects the competition between interactions and spin-orbit effects.

\subsection{ Low energy spectrum of symmetric double dot}

\begin{figure}[h]
\begin{center}
\includegraphics[scale=0.6,angle=0]{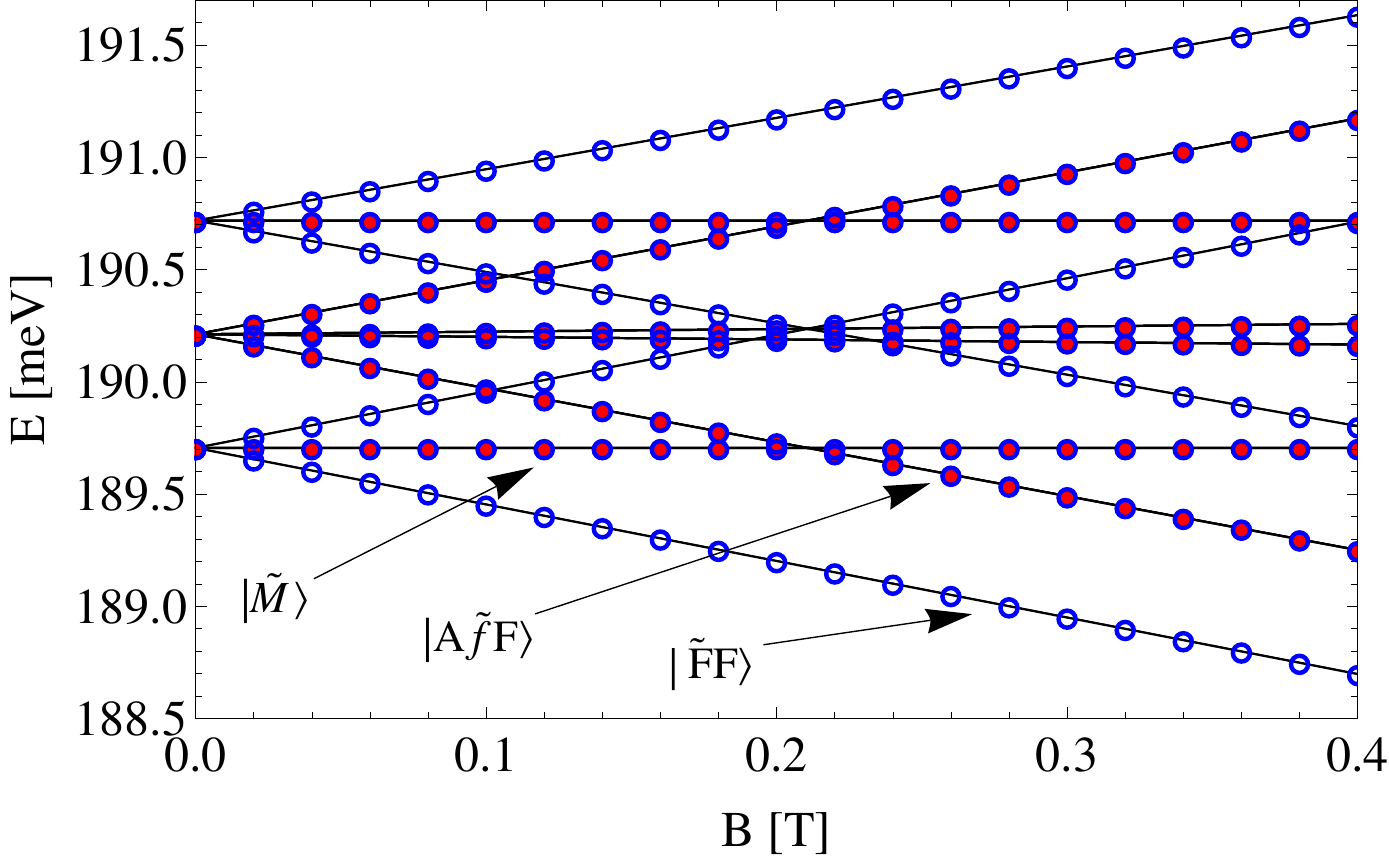}
\caption{(Color online) Low-energy spectrum of a double dot as a function of $B$. Solid symbols (red online) correspond to states belonging to the antisymmetric spin-valley multiplet and open symbols (blue online) to the symmetric spin-valley multiplet. The solid curves correspond to the effective Hamiltonian description.}
\label{ED0}
\end{center}
\end{figure}

First, we analyze the numerical results obtained with the multiband treatment. Figure~\ref{ED0} shows the low-energy spectrum as a function of $B$ for $k_d=2.5$ in a double dot with $\Delta=0$ (zero detuning).
In the low-energy spectrum, we can recognize the symmetric and antisymmetric spin-valley multiplets discussed in Table/Fig.~\ref{table1}.
The energy difference $\epsilon$ between the two multiplets is not observable in the energy range of  Fig.~\ref{ED0}.  Since interdot tunneling is very small compared with the interaction energy, the two electrons occupy different dots to avoid strong intradot interactions. The interdot interaction is almost independent of the orbitals occupied in each dot.
  Thus, at small detuning there is a negligible occupation of higher bands.

In Fig.~\ref{ED0},  we identify three states that will be relevant for the discussion of Pauli blockade. One of them has mixed valley-spin symmetry and we label it as $\ket{\tilde{M}}$. The second state is antiferromagnetic in spin and ferromagnetic in valley degree of freedom and we label it as $\ket{\tilde{AfF}}$. The third state, labeled $\ket{\tilde{FF}}$,  is ferromagnetic in both spin and valley degrees of freedom.
 States $\ket{\tilde{M}}$, $\ket{\tilde{AfF}}$ belong to the antisymmetric spin-valley multiplet and $\ket{\tilde{FF}}$ belongs the symmetric spin-valley multiplet. Their configurations are approximately
\begin{eqnarray}
\ket{\tilde{M}}&\approx & |1L,1R\rangle^{+}\ket{K \downarrow ; K' \uparrow }^-,\label{t1}\\
\ket{\tilde{AfF}}&\approx & |1L,1R\rangle^{+}\ket{K \downarrow ; K \uparrow }^-, \mbox{ and}\label{t2}\\
\ket{\tilde{FF}}&\approx & |1L,1R\rangle^{-}\ket{K \downarrow ; K \downarrow }.\label{t3}
\end{eqnarray}
 Using exact diagonalization, we extract the wave function and conclude that the states $|1L,1R\rangle^{\pm}$ are 
 given, to a very good approximation, by a single Slater determinant formed with left and right orbitals in the lowest band.

\begin{figure}[h]
\begin{center}
\includegraphics[scale=0.5,angle=0]{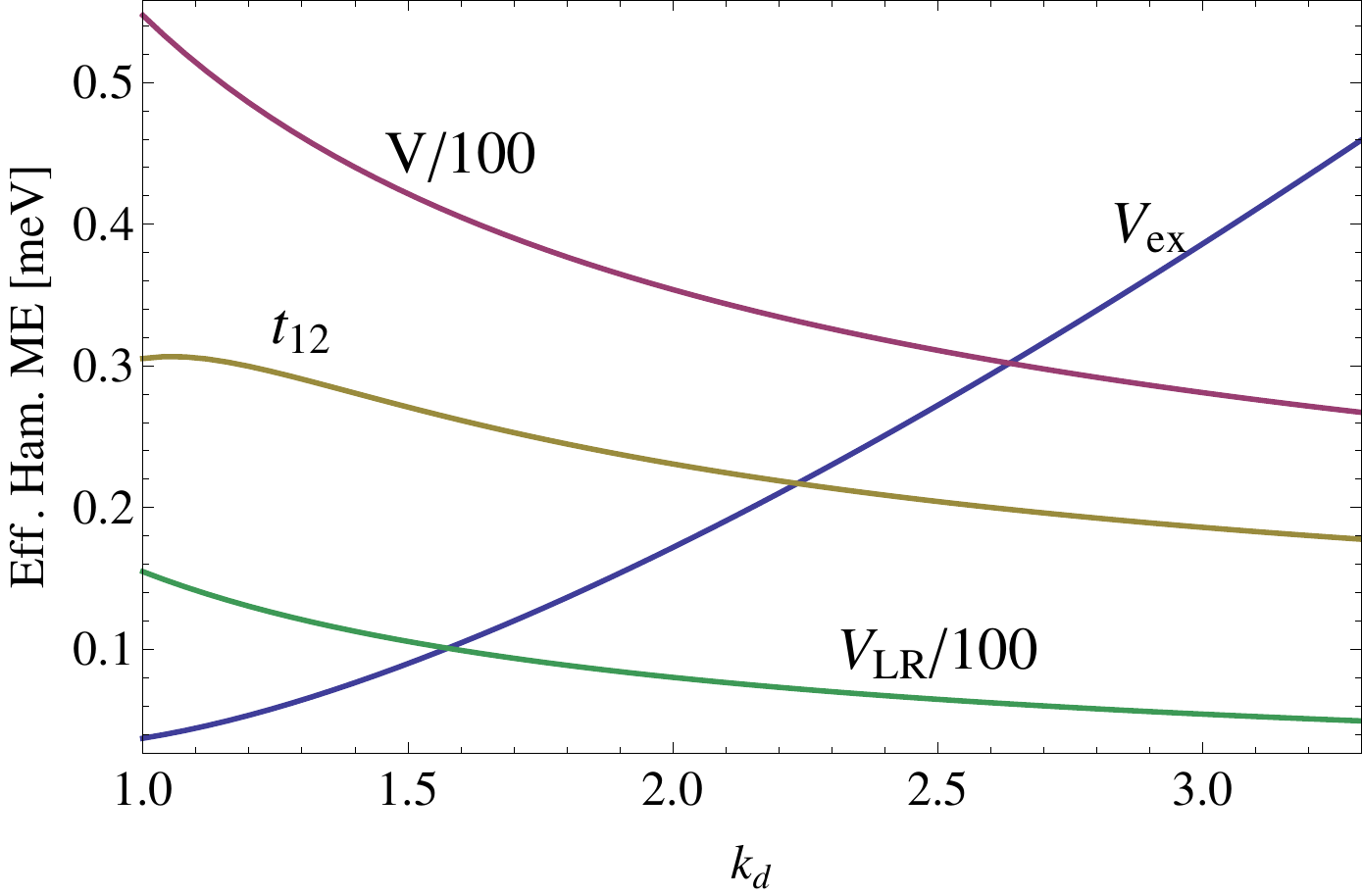}
\caption{(Color online) Parameters of the effective Hamiltonian as a function of the dielectric constant.}
\label{ParamEfHam}
\end{center}
\end{figure}

We can approximate the energies of $\ket{\tilde{M}}$, $\ket{\tilde{AfF}}$  and  $\ket{\tilde{FF}}$ using Eqs.~\eqref{E1pA2}, \eqref{E11ap1} and \eqref{E11ap2},
 \begin{eqnarray}
E_{\ket{\tilde{M}}}&\approx&2E_1^c(V_g)-2t_{12}^2/(V-V_{LR}- V_{ex})+V_{LR}\nonumber\\
&& -\Delta_{SO},\\
E_{\ket{\tilde{AfF}}}&\approx&2E_1^c(V_g)-2t_{12}^2/(V-V_{LR}- V_{ex})+V_{LR}\nonumber\\
&& -2 B \mu_{orb},\\
E_{\ket{\tilde{FF}}}&\approx&2E_1^c(V_g)-2t_{12}^2/(V-V_{LR}+V_{ex})+V_{LR} \nonumber\\
&&-\Delta_{SO} -2 B (\mu_{orb} + \mu_{spin}).\label{E3Tap}
\end{eqnarray}
At $B=0$, the  ground state $\ket{\tilde{M}}$ is only separated by the very small superexchange energy $\epsilon\approx  4 t_{12}^2 V_{ex}/(V-V_{LR})^2$ from the lowest spin-valley symmetric triplet that contains $\ket{\tilde{FF}}$. These four states are separated from the rest of the spectrum by a much larger energy scale given by $\Delta_{SO}$. This energy structure resembles the  singlet-triplet splitting in GaAs double quantum dots. At finite fields, $\ket{\tilde{FF}}$ is the ground state. The first excited state changes with increasing magnetic field from $ \ket{\tilde{M}}$ to $\ket{\tilde{AfF}}$ at $B_c= \Delta_{SO}/(2 \mu_{orb})$, as shown in Fig.~\ref{ED0}. This crossing between two antisymmetric spin-valley states has no analogous in standard GaAs quantum dots.

From the analysis of the single and double dot spectrum, we can obtain the parameters of the charge effective Hamiltonian (Eqs.~\ref{HmodelS},~\ref{HmodelAS}). Figure~\ref{ParamEfHam} presents the parameters $V$, $V_{LR}$, $t_{12}$ and $V_{ex}$ for a double dot with $L=70$~nm, $a=20$~nm, $V_g=78$~meV, $R=2.5$~nm, and $\Delta=0$. The $V$ and $V_{ex}$ are obtained from the two-electron spectrum in a single dot and $V_{LR}$ and $t_{12}$ obtained from double dot spectrum. The black solid curves in Fig.~\ref{ED0} show the prediction from the model using the parameters from Fig.~\ref{ParamEfHam}.

\subsection{Low energy spectrum for large detuning: Two electrons in a single dot}

 When the detuning becomes larger than  the intradot interaction, both electrons occupy the same dot, and the charge degree of freedom of the low-energy eigenstates can be described by the $|2L,0R\rangle$ configuration.
 In this regime, the energy spectrum resembles the one obtained for two electrons in an isolated dot\cite{wunsch09}.

\begin{figure}[h]
\begin{center}
\includegraphics[scale=0.55,angle=0]{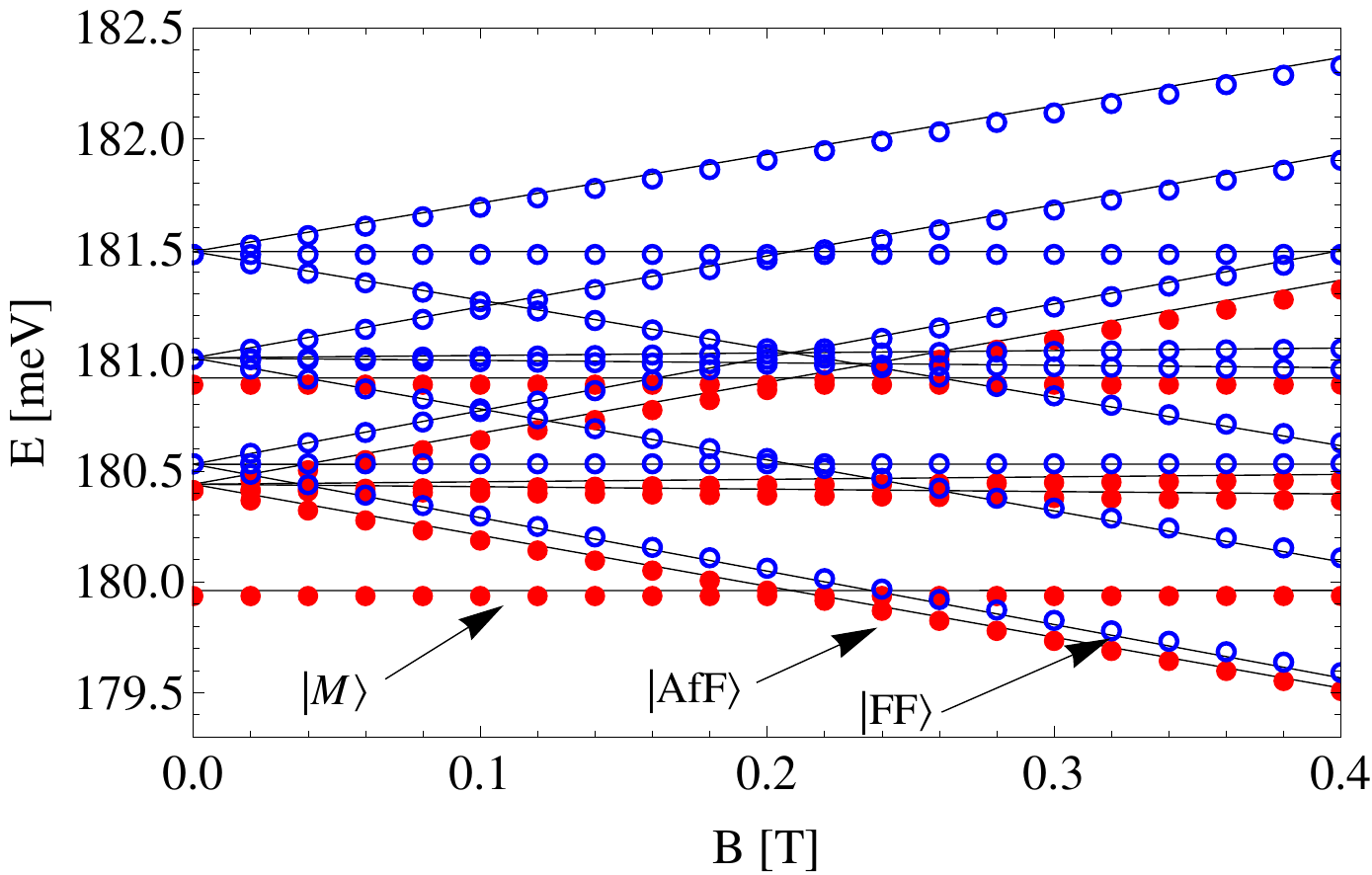}
\caption{(Color online)  Low energy spectrum  at large detuning ($\Delta\approx35$~meV). Solid symbols (red online) correspond to states belonging to the antisymmetric spin-valley multiplet, and open symbols (blue online) to the symmetric spin-valley multiplet. Curves represent the model predictions.}
\label{Ener2p1d}
\end{center}
\end{figure}

Figure~\ref{Ener2p1d} presents the low energy spectra for large detuning for $k_d=2.5$. The solid curves represent the effective Hamiltonian predictions. The multiband structure of the solutions introduce corrections to the spin-valley dependence of the spectrum which can be absorbed in effective $\mu_{orb}$ and $\Delta_{SO}$. However, this corrections are small and correspond to a few percent changes to the bare $\mu_{orb}$ and $\Delta_{SO}$ values.

At zero magnetic field, the  antisymmetric spin-valley multiplet is favored. This is in agreement with Lieb Mattis theorem\cite{Lieb61} that states that the two particle ground state always has a symmetric orbital part. In our two-electron system, we can understand this prediction from the analysis of the orbital symmetry of the wave function. In the noninteracting  limit, the orbital ground state is constructed  with both electrons in the lowest band corresponding to a symmetric orbital wave function, i.e., an antisymmetric spin-valley wave function. To form an antisymmetric orbital wave function at least one electron has to occupy an excited state.  If the electrons were noninteracting, $\epsilon$ would be given by the level splitting, $\hbar\omega_0$, between the first two bands.
 However, interactions substantially reduce the energy difference $\epsilon$ between the two multiplets as shown in Fig.~\ref{Ener2p1d}. Consequently, $\epsilon$ can be changed by tuning the ratio $U/\hbar \omega_0$, which can be achieved by changing the dielectric constant (modifying $U$) or changing the dot length (modifying both $U$ and $\omega_0$). In the limit of infinite interactions, the electrons are strongly localized at the positions that minimize the interaction energy. Then the orbital symmetry of the wave function becomes irrelevant and $\epsilon$ vanishes. This effect is a signature of a formation of a Wigner molecule\cite{wunsch09,secchi2009B,secchi2009cvs}.

In the effective Hamiltonian, the formation of a Wigner molecule is manifested in the reduction of the parameter $V_{ex}$. This is evident in Fig.~\ref{ParamEfHam} that shows that $V_{ex}$ is a growing function of the dielectric constant. Even though the results of Fig.~\ref{ParamEfHam} are for $\Delta=0$, we note that $V$ and $V_{ex}$ depend weakly on the detuning and can be approximated by $\Delta=0$ predictions for all detunings studied here $\Delta< 35$ meV. In contrast, the tunneling $t_{12}$ is strongly affected by the detuning and is reduced almost by approximately a factor of two in comparison with the $\Delta=0$ case.

The reduction of $\epsilon$  implies that the exact eigenstates become strongly correlated and cannot be written as noninteracting wave functions to characterize them. However, we can still label the low-lying $(2,0)$ states according to their conserved quantum numbers:
\begin{eqnarray}
\ket{M}&=&    |2L,0R\rangle^+  \ket{K \downarrow ; K' \uparrow }^-,\label{eqM}\\
\ket{AfF}&=&      |2L,0R\rangle^+  \ket{K \downarrow ; K \uparrow }^-, \mbox{ and}\\
\ket{FF}&=&   |2L,0R\rangle^- \ket{K \downarrow ; K \downarrow }\label{eqFF}
\end{eqnarray}
Here the states $|2L,0R\rangle^\pm$  are correlated orbital states of two electrons in the left dot.
Our numerical calculations shows that several bands are needed to represent these states accurately.
 We note that Secchi and Rontani found qualitatively the same correlation effects for a parabolic well with weak confinement~\cite{secchi2009cvs}, which suggests the robustness of these correlation effects.
 We note that because of the conserved symmetries, the eigenstates at  small detuning $\ket{\tilde{M}}$, $\ket{\tilde{AfF}}$, and $\ket{\tilde{FF}}$  will evolve in $\ket{M}$, $\ket{AfF}$, and $\ket{FF}$  for large detuning.

Using the effective Hamiltonian along with the approximate description of the single particle energies [Eqs.\ref{E1pA2},
\ref{E20ap1} and \ref{E20ap2}], we can obtain simple expression for the (2,0) configuration energies:
\begin{eqnarray}
E_{\ket{M}}&\approx&2E_1^c(V_g)+V-V_{ex}-\Delta_{SO}-\Delta, \\
E_{\ket{AfF}}&\approx&2E_1^c(V_g)+V-V_{ex}-2 B \mu_{orb}-\Delta, \,\,\mbox{and}\\
E_{\ket{FF}}&\approx&2E_1^c(V_g)+V+V_{ex}-\Delta_{SO}\nonumber\\
& &-2 B (\mu_{orb} + \mu_{spin})-\Delta.\label{E3ap}
\end{eqnarray}
At $B=0$,  $\ket{M}$ is the ground state. Above a critical magnetic field, there is a ground-state transition to either $\ket{AfF}$ ( $\Delta_{SO}<\epsilon$) or $\ket{FF}$ ( $\Delta_{SO}>\epsilon$) due to the orbital Zeeman term.
Thus, the reduction of $\epsilon$ leads to a ground state transition to the ferromagnetic state $\ket{FF}$ at finite magnetic field. In practice, the formation of a ferromagnetic ground state can be experimentally controlled by changing the length of the dots.

\subsection{Transition from a double- to a single-dot regime.}

We now analyze the transition between the limiting behaviors discussed in the previous two sections.
\begin{figure}[h]
\begin{center}
\includegraphics[scale=0.5,angle=0]{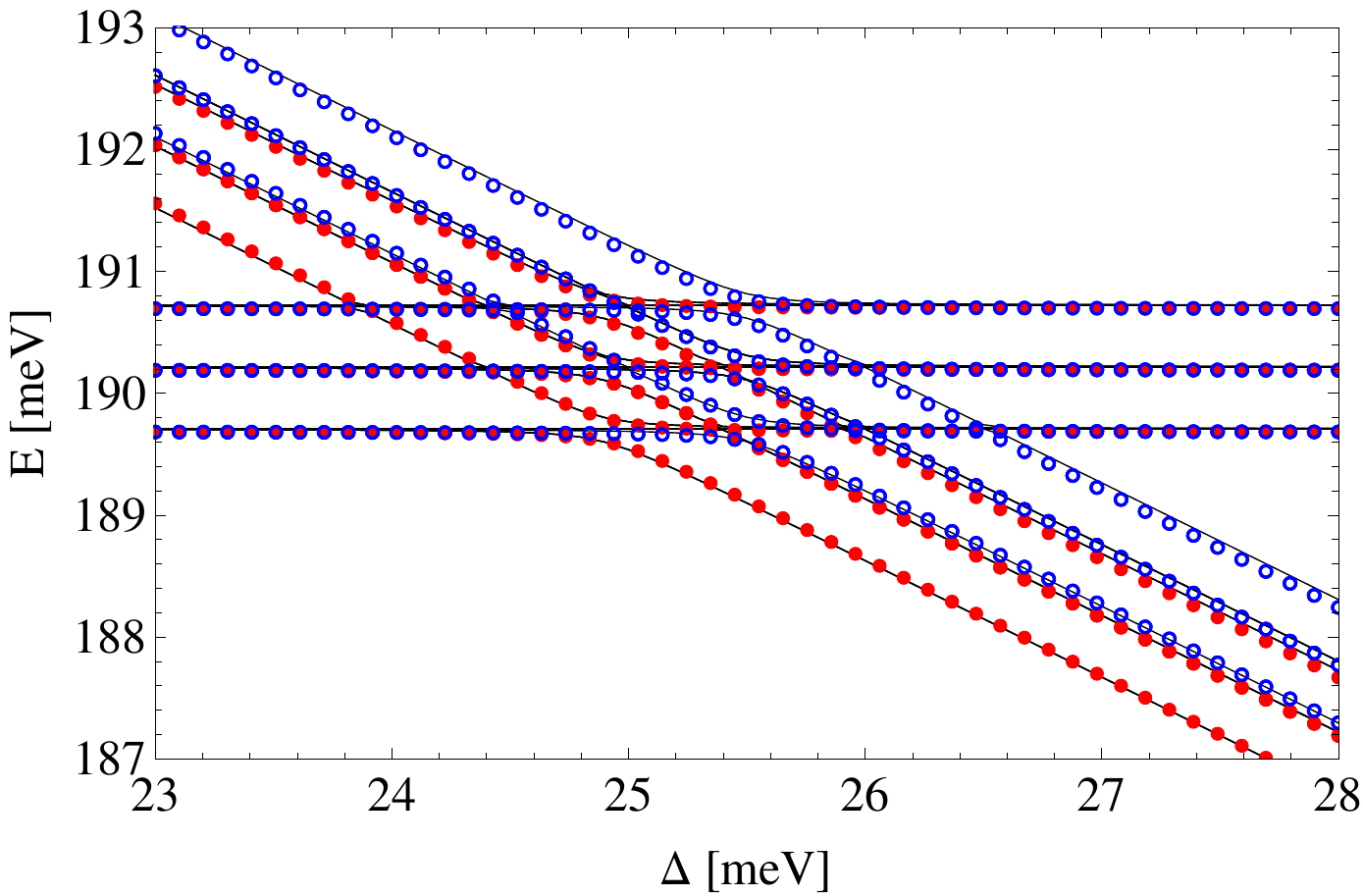}
\caption{(Color online) Zoom of first crossing of (1,1) and (2,0) states for $B=0$. Solid symbols (red online) correspond to states belonging to the antisymmetric spin-valley multiplet, and open symbols (blue online) to the symmetric spin-valley multiplet. Curves represent the model predictions. }
\label{Ezoom1}
\end{center}
\end{figure}

Figure~\ref{Ezoom1} shows the crossover between $|1L,1R\rangle$ states and $|2L,0R\rangle$ states of the double dot system at $B=0$.
In absence of interdot tunneling, the two-electron spectrum shows sharp crossings between the  $|1L,1R\rangle$ states and $|2L,0R\rangle$ states with increasing detuning.
Because of the interdot tunneling, crossings between states with the same symmetries turn into avoided crossings. The avoided crossings occur at the same critical detuning for states of the same multiplet. The avoided crossings within the multiplet with an antisymmetric orbital part occur at relatively larger detuning since the tunneling electron is forced to occupy an excited band.

The crossover regime, presented in Fig.~\ref{Ezoom1}, is strongly affected by interactions.
The difference between the critical detunings belonging to the avoided crossings of states with symmetric and antisymmetric orbital part is a direct measure for the energy splitting $\epsilon$.
 Furthermore, correlations in the  $|2L,0R\rangle$ states decrease the tunneling coupling to the corresponding $|1L,1R\rangle$ state, leading to a sharper avoided crossing.

\begin{figure}[h]
\begin{center}
\includegraphics[scale=0.75,angle=0]{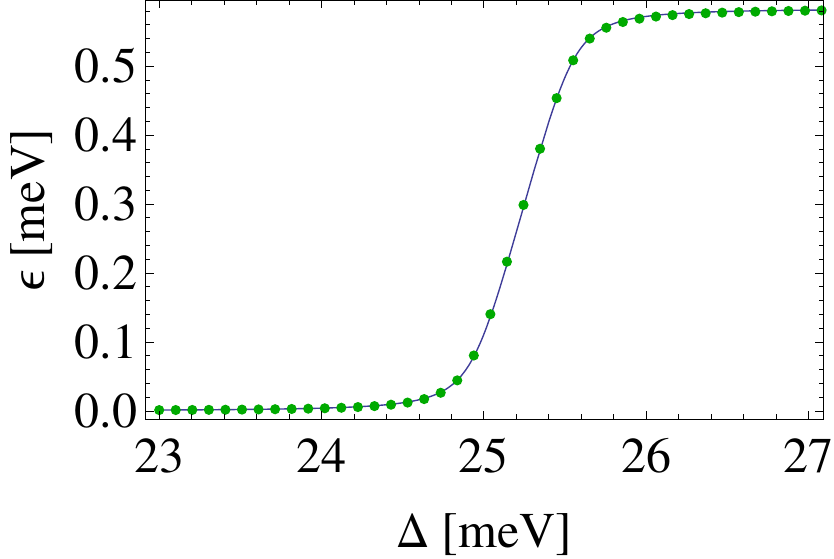}
\caption{(Color online) $\epsilon$ as a function of detuning, symbols correspond to the numerical results and the curve is the effective Hamiltonian prediction.}
\label{epsilon}
\end{center}
\end{figure}
In the effective Hamiltonian description, the lowest $|1L,1R\rangle$ and $|2L,0R\rangle$ configurations are close in energy when $\Delta\sim V-V_{LR}$. In this regime, the $|0L,2R\rangle$ configurations are energetically suppressed and do not affect the low energy spectrum. Within this approximation, we can obtain a simple expression for $\epsilon$,
 \begin{eqnarray}
 \epsilon=V_{ex}+\frac{1}{2}\left(\sqrt{4t_{12}^2+(\Delta-V_{ex}-V+V_{LR})^2}\right.\nonumber\\
\left.-\sqrt{4t_{12}^2+(\Delta+V_{ex}-V+V_{LR})^2}\right)
 \end{eqnarray}
This expression compares well with the numerical prediction as shown in Fig.~\ref{epsilon}.

\begin{figure}[h]
\begin{center}
\includegraphics[scale=0.12,angle=0]{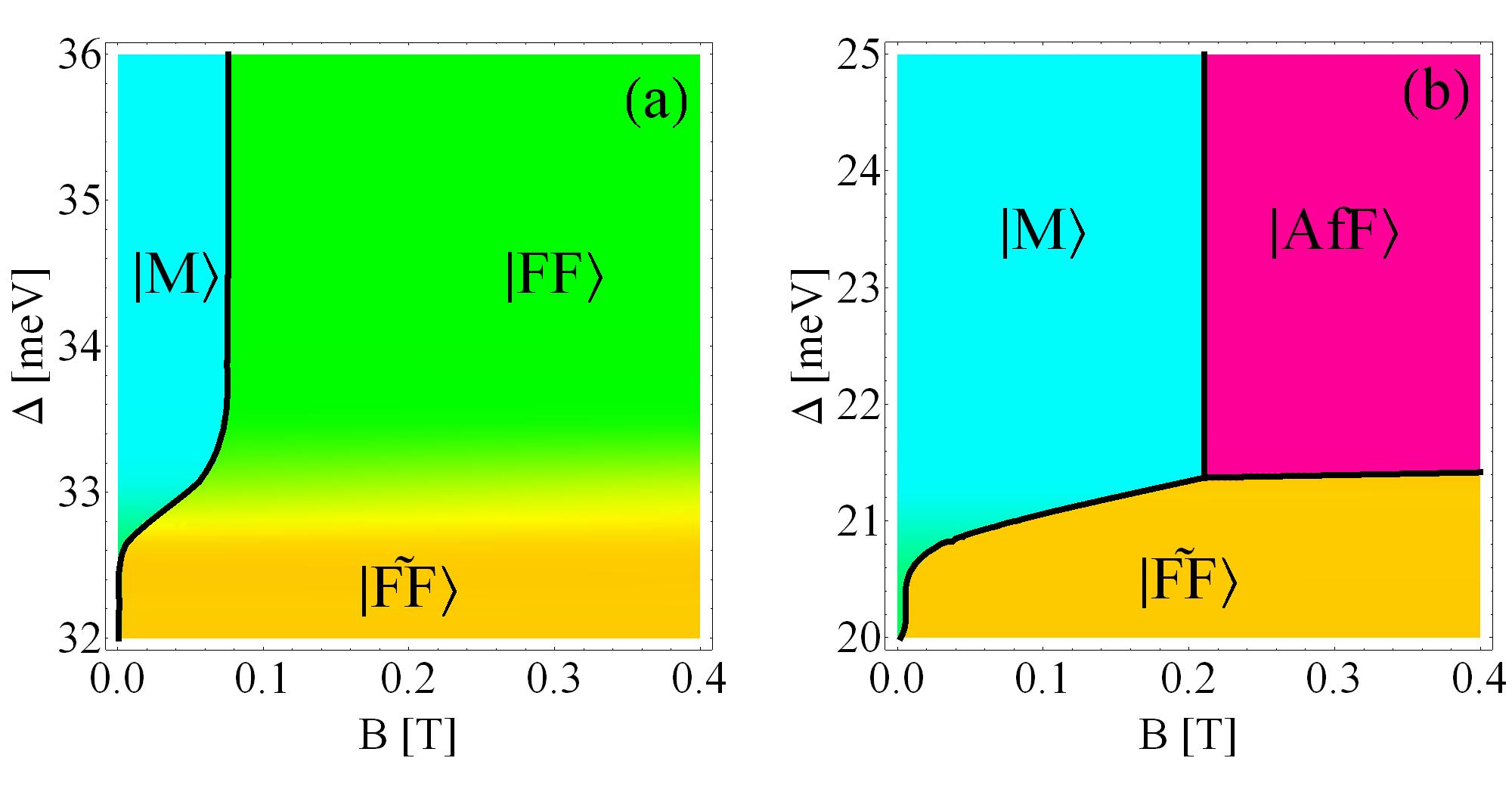}
\caption{(Color online) Phase diagram for a double quantum dot with five bands and  $k_d=1.5$ (a) and $k_d=3.5$ (b). Black solid lines separate regions with different spin-valley symmetry. Regions with the same greyscale ( color online) characterize the ground state.}
\label{DP5b}
\end{center}
\end{figure}

From the analysis of the low-energy spectrum, e. g., Fig.~\ref{Ezoom1}, we can extract
the phase diagram as a function of the detuning $\Delta$ and the magnetic field $B$. The phase diagrams for strong ($k_d=1.5$) and weak ($k_d=3.5$) interactions, presented in Fig.~\ref{DP5b}, exhibit clear differences.
At exactly zero magnetic field, the ground state always has a symmetric orbital wave function ($\ket{M}$ and $\ket{\tilde{M}}$).
This state is preferred with respect to other states within the same multiplet by the spin-orbit coupling $\Delta_{SO}$ and, with respect to states of the other multiplet, by the energy gap $\epsilon$. The ground state changes at a critical magnetic field to a valley-polarized state. The valley polarized ground state has symmetric orbital part for $\Delta_{SO}<\epsilon$ and an antisymmetric orbital part for $\Delta_{SO}>\epsilon$.
For small detuning $\Delta$, the energy splitting $\epsilon$ is caused by the small superexchange and $\epsilon<\Delta_{SO}$. However, for large detuning, both electrons are on the same dot, and the value of  $\epsilon$ can be larger or smaller than $\Delta_{SO}$. Thus, if $2 V_{ex}>\Delta_{SO}$, there is a transition to a valley-polarized state with a symmetric orbital part, $\ket{AfF}$, as detuning is increased [as observed in Fig.~\ref{DP5b}~(b)].

\subsection{Sequential transport}
\label{subsecST}

\begin{figure}[h]
\begin{center}
\includegraphics[width=\linewidth,angle=0]{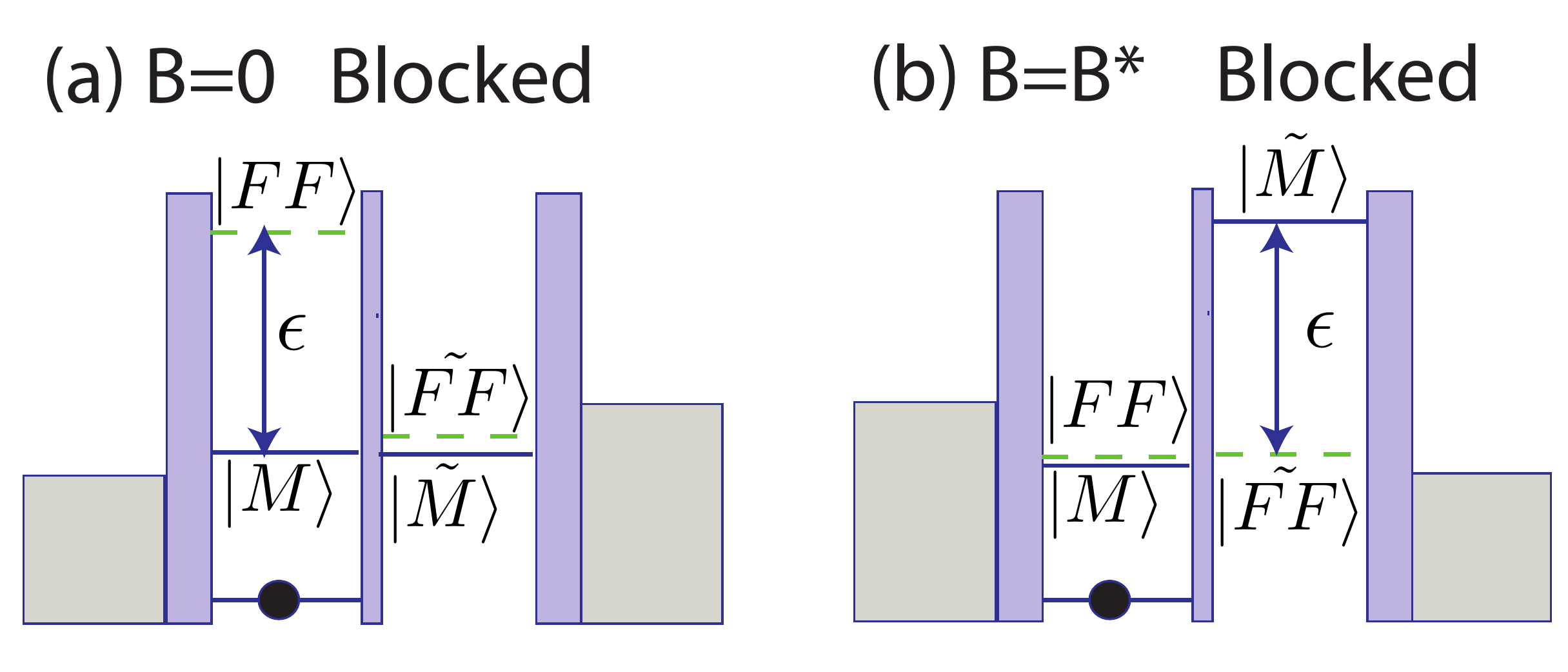}
\caption{ (Color online) Schematic representation of a Pauli blockade for: (a) zero magnetic field and negative bias, and (b) finite magnetic field and positive bias. The white areas in the center of each figure correspond to the two dots. The black circle in the left dot stands for the position and energy of one of the electrons. The states $\ket{M}$ and $\ket{\tilde{M}}$ are represented by solid lines, and the states $\ket{FF}$ and $\ket{\tilde{FF}}$ by dashed lines. These states are organized vertically according to their energies, and also represent the possible positions of the second electron. The wide gray rectangles next to the left and right dots correspond to the energy bias.}
\label{Blockade}
\end{center}
\end{figure}

In this section, we discuss how the two-electron eigenspectrum affects sequential transport through a double dot. In particular, we analyze the existence of Pauli blockade in serially coupled dots that would lead to a current rectification in DC transport\cite{Ono2002,petta2005cmc}. Figure~\ref{Blockade} shows a schematic description of two situations when a Pauli blockade occurs.
In both cases, the double-dot system is detuned such that there is always at least one electron in the left dot.
States relevant for transport through the double dot are the $|2L,0R\rangle$  states $\ket{M}$ and $\ket{FF}$ as well as the $|1L,1R\rangle$ states $\ket{\tilde{M}}$ and $\ket{\tilde{FF}}$. In Fig.~\ref{Blockade}, the vertical axes denotes energy and the horizontal axes denotes the spatial coordinate along the nanotube. The center of each figure shows the double dot created by tunnel barriers to the contacts and between the dots. The gray rectangles to the left and right of the double dot are the Fermi seas in the contacts.

For a positive bias voltage (opposite to Fig.~\ref{Blockade} (a)) , the electrochemical potential in the left contact is larger than in the right one, and current flows via a sequential tunneling process $|1L,0R\rangle\to|2L,0R\rangle\to|1L,1R\rangle\to|1L,0R\rangle$. Interdot tunneling is assumed to conserve the spin-valley degree of freedom and allows for transitions between states $\ket{M}$ and $\ket{\tilde{M}}$ or $\ket{FF}$ and $\ket{\tilde{FF}}$. Because of the sequential transport setup, the left dot couples to the left reservoir, allowing for transitions between the $|1L,0R\rangle$ and $|2L,0R\rangle$ states. Analogously, the right contact allows for transitions between the $|1L,0R\rangle$ and $|1L,1R\rangle$ states. Figure~\ref{Blockade} (a) considers Pauli blockade at $B=0$.

 In this scenario, the current is blocked when the state $\ket{\tilde{FF}}$ is occupied by an electron tunneling in from the right reservoir.
Once in state $\ket{\tilde{FF}}$, the electron cannot tunnel back because of the filled Fermi sea in the right contact.
Also, if $\epsilon$ is large, the electron cannot tunnel to the left dot since a transition to the  $\ket{FF}$ state is energetically suppressed [see Fig.~\ref{Blockade} (a)]. However, $\epsilon$ can be strongly reduced, allowing for a finite exit rate from state $\ket{\tilde{FF}}$.
This effect might explain the absence of a Pauli blockade in Ref.~\onlinecite{Gotz09}.  This Pauli blockade implies a rectification of the current since, by inverting the bias voltage, a finite current can flow\cite{footnote1}.

In order to make these statements more quantitatively we calculate the stationary current with a rate equation approach \cite{Stoof96,sprekeler04,wunsch05}. We describe the regime of possible blockade depicted in Fig.\ref{Blockade} (a) while assuming spin-orbit coupling to be much bigger than temperature, transport voltage and external coupling the the reservoirs.

The rate equations and the resulting stationary current are given in Appendix~\ref{rate}.
The dependence of the stationary current $I_{bl}$ in the blockade setup and the probability to be in any of the three degenerate $|FF\rangle$ states are given by:
\begin{eqnarray}
I_{bl}&=&\frac{t_S^2 A}{B+C(2t_S^2+\epsilon^2)}\label{current1}\\
P_{\tilde{FF}}&=&\frac{ C(t_S^2+\epsilon^2)+D}{B+C(2t_S^2+\epsilon^2)} \end{eqnarray}
Here $\epsilon$ denotes the energy splitting between the states $\ket{M}$ and $\ket{FF}$ and $t_S$ is the interdot tunneling rate between states $\ket{FF}$
 and $\ket{\tilde{FF}}$. The constants $A,B,C,D$ depend on the coupling strength to the contacts and spectral weights for the tunneling probabilities and their form is given in Appendix \ref{rate}.

If $\epsilon$ is the dominant energy scale then the current is suppressed as $I_{bl}\propto 1/\epsilon^2$ and the double dot gets stuck in states $|\tilde{FF}\rangle$ with $P_{\tilde{FF}}\to 1$. This is the regime of Pauli-blockade. However, we find that due to interaction effects interdot tunneling $t_S$ can even exceed the energy splitting $\epsilon$, thus removing the blockade mechanism.
\begin{figure}[h]
\begin{center}
\includegraphics[width=0.7\linewidth,angle=0]{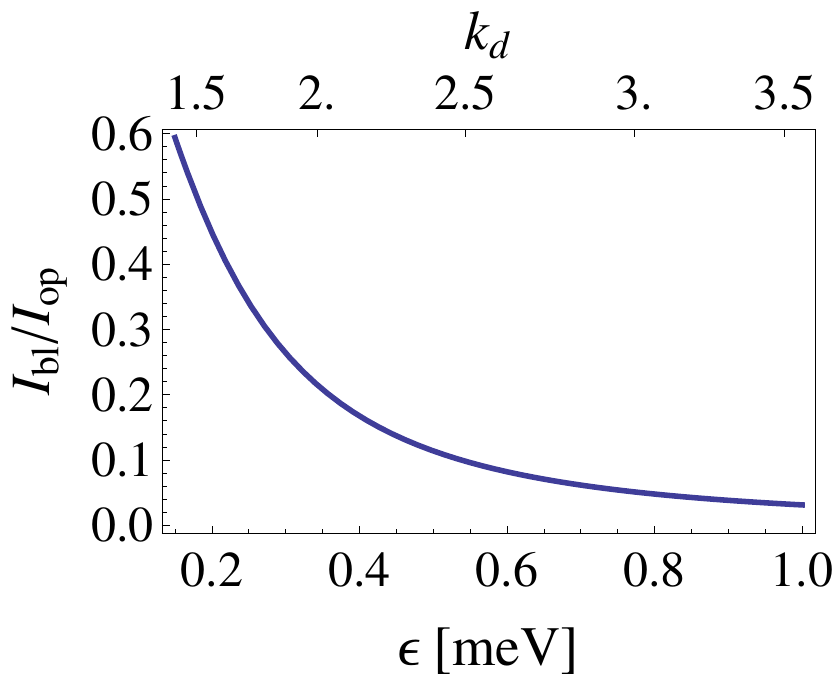}
\caption{(Color online) Ratio between leakage current $I_{bl}$~\eqref{current1} corresponding to setup in Fig.~\ref{Blockade} (a) and  the finite current for reversed transport voltage $I_{op}$~\eqref{current2}, i.e., the current in the open (unblocked) direction. Used parameters $\Gamma_L=\Gamma_R=0.01$meV, $t_{AS}=t_S=0.1$meV and $S_M=0.7$ $S_F=1$. }
\label{Blockade2}
\end{center}
\end{figure}

Due to Pauli-blockade the double dot acts as a current rectifier, since by reversing the transport voltage in Fig.~\ref{Blockade} a) a finite current can flow.
Fig.~\ref{Blockade2} illustrates how effective the double dot acts as current rectifier as function of $\epsilon$.

At finite magnetic field, the transport behavior can also be very interesting.
For example, if interactions are strong enough to suppress $\epsilon$ to become on the order of the spin-orbit coupling ($\Delta_{SO}\geq\epsilon$) but still much larger than the interdot tunneling ($\epsilon\gg t$), then a current blockade can occur in the opposite bias direction of the zero field case. This situation, depicted in Fig.~\ref{Blockade} (b),  is achieved by changing both the magnetic field and the detuning.  A magnetic field $B^*\approx \epsilon/2(\mu_{orb}+\mu_{spin})$ is applied so that states $\ket{FF}$ and $\ket{M}$ are degenerate, while
in the $|1L,1R\rangle$  charge configuration, the state $\ket{\tilde{FF}}$ is ground state and separated from state $\ket{\tilde{M}}$ by the energy $\epsilon$ (because of the applied field $B^*$). The physical situation of Fig.~\ref{Blockade} (b) resembles that of Fig.~\ref{Blockade} (a), and also results in a current blockade.  This ability to control the direction of the current rectification by varying the magnetic field and detuning can have applications to carbon nanotube-based spintronic proposals.

\subsection{1D-Quantum dot arrays}
\label{Chain}

The rich physics associated with the strong correlations of double occupied dots is expected to have significant impact in the many-body behavior of a one-dimensional chain of dots.
The double dot analysis carried out here can be used as a starting point to study the behavior of a linear array  of coupled dots.  According to our analysis, there exists an interaction regime in which the behavior of two electrons in the same dot is strongly correlated and controlled by $\epsilon$ while the behavior of electrons in different dots is weakly correlated and can be accurately described considering only the lowest orbital of each dot.
 In order to generate an effective Hamiltonian that captures the essential effects of strong onsite correlations, we apply the ideas of Hubbard operators  to describe double occupied dots or doublons (see i.e. Ref.~\onlinecite{ovchinnikov2004hubbard}). This description assumes that the strong onsite Coulomb repulsion suppresses the probability to occupy a dot with more than two electrons. This is a good approximation for low enough fillings.

Formally, we start from the complete Hamiltonian that describes a chain of dots.
This Hamiltonian represents an extension of the double dot Hamiltonian introduced in the Appendix.

We introduce the operators $a^\dagger_{r,n,\alpha}$ which create an electron in the $r$ dot, in the $n$ orbital state and with a given spin-valley configuration $\alpha$. To simplify the notation we introduce $a^\dagger_{r,\alpha}\equiv a^\dagger_{r,1,\alpha}$ for electrons in the lowest band which describe well the singly occupied dots.
 The state that describes two electrons in the $r$ dot can be expanded in single particle basis as
\begin{eqnarray}
|d_{\eta,r, \alpha,\alpha'}\rangle&=&f_{\alpha,\alpha'}^{-1}\sum_{n\le m}\beta_{\eta,n,m}(a^\dagger_{r,n,\alpha}a^\dagger_{r,m,\alpha'}\notag\\
& &+(-1)^\eta a^\dagger_{r,n,\alpha'}a^\dagger_{r,m,\alpha})|0\rangle
\label{ddef}
\end{eqnarray}
Here, $\eta$ labels the spin-valley symmetry of the two-electron state $\eta=1$ for antisymmetric spin-valley states and $\eta=2$ for symmetric spin-valley states and $f_{\alpha,\alpha'}=\sqrt{1+\delta_{\alpha,\alpha'}}$.
Equation~\ref{ddef} defines the doublon operator such that $|d_{\eta,r,\alpha,\alpha'}\rangle=d^\dagger_{\eta,r,\alpha,\alpha'}|0\rangle$ and $d^\dagger_{\eta,r,\alpha,\alpha'}=(-1)^\eta d^\dagger_{\eta,r,\alpha',\alpha}$.
Note that the $\beta_{\eta,n,m}$ do not depend on $\alpha$ and $\alpha'$. 
The coefficients $\beta_{\eta,n,m}$ can be obtained by diagonalizing the local part of the Hamiltonian which amounts to solve two electrons in a single dot. We now assume that the occupation of higher excited two-particle states, which are separated by an energy gap of the order of the single particle level spacing, can be neglected. This should be a good approximation for the low energy spectrum.

The effective Hamiltonian is obtained by projecting the complete Hamiltonian onto the subspace of empty, singly and doubly occupied dots and can be written as
\begin{equation}
H_{eff}=P(H_e+H_d+H_{ed}+V_c)P.
\label{Hchain}
\end{equation}
where $P\equiv \prod_i P_i$ represent a projector to the physically allowed subspace where $P_i=|0\rangle_i\langle0|+\sum_\alpha|\alpha\rangle_i\langle\alpha|+\sum_\eta|d_\eta\rangle_i\langle d_\eta|$ (see Ref.~\onlinecite{duan2005effective}).
 $H_e$ describes the behavior of single occupied dots, $H_d$ describes the behavior of double occupied, $H_{ed}$ contains a coupling between singly and doubly occupied dots and $V_c$ represents the Coulomb interaction between different sites. The explicit form of these contribution is:
\begin{eqnarray}
H_{e}&=&\sum_{r,\alpha } E_{r,\alpha}a_{ r \alpha }^{\dagger}a^{}_{r\alpha}+\sum_{\langle r,r'\rangle,\alpha} t_{r,r'} a_{ r \alpha }^{\dagger}a^{}_{r'\alpha}\\
H_{d}&=&\sum_{r,\eta, \alpha\leq\alpha'}E^{2b,\eta}_{r,\alpha,\alpha'}d_{\eta, r \alpha\alpha' }^{\dagger}d^{}_{\eta,r\alpha\alpha'}\nonumber\\
&&+\sum_{\langle r,r'\rangle,\eta, \alpha\leq\alpha'} t^{(d)}_{\eta,r,r'}  d_{\eta, r \alpha\alpha' }^{\dagger}d^{}_{\eta, r'\alpha\alpha'}\\
H_{ed}&=& \sum_{\begin{subarray}{l}\langle r,r'\rangle,\eta,\eta',\\
\alpha_1,\alpha_2,\alpha_3\end{subarray}} t^{\eta,\eta'}_{r,r'} f_{\alpha_1\alpha_3}f_{\alpha_2\alpha_3} d_{\eta r \alpha_1 \alpha_3}^\dag a_{r' \alpha_2}^\dag a_{r\alpha_1}d_{\eta'r'\alpha_2\alpha_3}\notag\\
&&+\sum_{\langle r,r'\rangle\eta,\alpha,\alpha'} f_{\alpha\alpha'} g_{\eta,r,r'} [d^\dagger_{\eta,r,\alpha\alpha'}a_{r,\alpha} a_{r',\alpha'}+\mbox{H.c}]\nonumber\\
V_c&=&U_c\sum_{r\ne r'} \frac{n_r n_{r'}}{r-r'}
\end{eqnarray}
The diagonal part of $H_{e}$ and $H_{d}$ represent the electron and doublons energies composed of the spin-valley energy and onsite energies ($E_{r,\alpha}=E^c_1(V_r)+E_\alpha$ and  $E^{2b,\eta}_{r,\alpha,\alpha'}=2E^c_1(V_r)+E_\alpha+E_{\alpha'}+V+(-1)^\eta V_{ex}$) , and $t$ and $t^{(d)}$ the tunnelings of the electrons and doublons. The $H_{ed}$ contains a term that describes the hopping of an electron from the doubly occupied site to a singly occupied site which in this language corresponds to an exchange of a doublon and an electron. The second term in  $H_{ed}$ represents the hopping of an electron from a singly occupied site to a singly occupied site and viceversa, which is represented as destroying two atoms and creating a doublon. Finally, we have the offsite Coulomb interaction $V_c$ in terms of the dot density $n_r$ defined as
\begin{equation}
n_r=\sum_\alpha a_{ r \alpha }^{\dagger}a^{}_{r\alpha}+2\sum_{\eta,\{\alpha,\alpha'\}}d_{\eta, r \alpha\alpha' }^{\dagger}d^{}_{\eta,r \alpha\alpha'}.
\end{equation}
Here, the summation $\{\alpha,\alpha'\}$ is restricted to $\alpha<\alpha'$ for $\eta=1$ and $\alpha\le\alpha'$ for $\eta=2$.
For the offsite interaction, we assume exclusively capacitive coupling and neglect a dependence on the symmetry of the two particle states.

Explicit expressions for the parameters $g_{\eta,r,r'}$, and $t^{\eta,\eta'}_{r,r'}$ can be obtained by comparing matrix elements of the exact and effective Hamiltonians. If the doublon solution is expressed as in Eq.~\ref{ddef}, the $g_{\eta,r,r'}$, and $t^{\eta,\eta'}_{r,r'}$ can be expanded in terms of the doublon expansion coefficient $\beta_{\eta,n,m}$ and the many-body Hamiltonian matrix elements (see e.g. Ref.~\onlinecite{duan2005effective}). Alternatively, these parameters can be obtained from comparison between exact and effective Hamiltonian solutions for two and three-electrons systems. These few-electron calculations might be challenging but allow the determination of the parameters needed for many-body calculations. The extraction of these coupling parameters is beyond the scope of the current article.

In the most general case the parameters that describe the effective Hamiltonian depend on lattice sites positions and can be controlled by changing the detuning in each lattice site $\Delta_{r}$. For example, an enhancement of the superexchange interactions can be achieved by detuning some of the dots and, therefore, reducing the energy cost of double occupancies.

The effective Hamiltonian can be applied to describe an array of coupled dots in many different regimes and its phase diagram might exhibit novel phases.
In particular, the existence of spin-valley ``triplet'' close in energy to the spin-valley ``singlet'' can lead to phenomena richer in comparison to the standard single-band Hubbard model \cite{Lieb68}. For example, for one particle per site (filling $n=1$),  it is known that the ground state of the usual Hubbard model has infinite susceptibility to spin dimerization\cite{Wilkens01}, i.e. formation of singlet/triplet bonds between nearest-neighbor sites.
However, the  ground state is not dimerized since the formation of singlet bonds,  say at sites  (2 i, 2 i+1), is penalized by the large energy cost  of   the remainder  triplet components  between sites  (2 i+1, 2 i+2).
In carbon nanotubes operating  in the strongly interacting regime , $\epsilon \ll 1$, even  spin-valley ``triplet'' states can lower their energy by virtual hoping. The later situation reduces the energy cost of triplet formation and the infinite susceptibilities to spin dimerization might in this case translate in actual  dimerization  of  the ground state.



Away from $n=1$, the presence of a spin-valley ``triplet'' can also significantly affect the magnetic structure of the system. In particular, for systems which already exhibit ferromagnetism in the standard single-band Hubbard model, the inclusion of the spin-valley ``triplet'' can strengthen and extend the ferromagnetic phase.
For example, generic 2 and 3 dimensional square lattice geometries, and others with similar connectivity conditions, exhibit  Nagaoka ferromagnetism\cite{Nagaoka1966}  when the gain of kinetic energy of a single hole exceeds the  decrease  of superexchange energy. The latter condition is fulfilled  at very large on-site interactions. In quantum dots in carbon nanotubes with small $\epsilon$, Nagaoka-type ferromagnetism might become stable at reduced value of the interaction
since the virtual hopping of the spin-valley triplets  reduces the super-exchange penalty of  having a polarized state.

For arrays with
filling $1<n<2$ and zero detuning,  the low energy physics consists of doubly and singly occupied dots while the occurrence of empty sites is strongly suppressed since it implies an increase of double occupancies. This implies that the number of electrons and doublons will be independently conserved and that the second term of $H_{ed}$ can be neglected. Also, the tunneling of the electrons and the doublons can be neglected.  Thus, the only relevant terms in the effective Hamiltonian are the onsite energies, the long-range Coulomb and the electron-doublon exchange (first term in $H_{ed}$). 
 This regime can lead to interesting phenomena when the electron-doublon exchange becomes comparable the multiplet splitting $\epsilon$.

Finally, it should be pointed out that the long-range Coulomb interaction can have a preponderant influence in the charge distribution in small arrays of quantum dots.

\section{Summary and conclusions}

We have presented a detailed study of the few-electron eigenspectrum of a double quantum dot in a semiconducting carbon nanotube.
We showed how the spin-valley physics leads to the formation of multiplets. The internal energy structure of the multiplet is practically unaffected by a change in either the confinement potential or the interaction strength, but the energy gap between different multiplets is strongly modified by both.  We showed that for sufficiently strong interactions, the spin-orbit coupling can exceed the energy splitting between states with symmetric and antisymmetric orbital parts for any detuning between the dots.
This situation modifies the two particle phase diagram. Above a critical interaction strength, the ground state at small, finite magnetic fields is always ferromagnetic  independent of the detuning. Furthermore, in this strongly interacting regime, the blockade of linear transport gradually disappears
since the reduction of the multiplets' energy splitting $\epsilon$ allows a finite tunneling probability from the  $|1L,1R\rangle$  state to the $|2L,0R\rangle$  state even for a symmetric spin-valley part.  Pauli-blockade physics will occur for weak enough Coulomb correlations, which can be suppressed by either working with short dots, or by covering the nanotube by strong dielectrics\cite{wunsch09}. We note that a well developed  Pauli-blockade is the precondition for realization of spin-qubits in double dots allowing for coherent singlet-triplet manipulation \cite{Hanson2007,Churchill09b}.

Our understanding of the double dot physics can be used as an starting point to analyze the behavior of an array  of coupled dots. The effective Hamiltonian, as presented in Eq.~\ref{Hchain}, can be applied to describe an array of coupled dots in certain regimes. In the future, we would like to explore the many-body physics associated with the strong onsite correlations and its consequences in the magnetic ordering of the system.

In the analysis presented here, we neglected terms that flip either the spin or valley degree of freedom, or both.  However, such mechanisms could change the various level crossings to avoided crossings and thus open new ways to control the spin-valley degree of freedom as well as the transport properties of this system\cite{murgida2007coherent,murgida2009coherent,Burkard09,Burkard10}.

\section{Acknowledgments}
We thank F. Kuemmeth, H. Churchill, and H. Bluhm
 for illuminating discussions. B. W. is funded by the German Science Foundation under grant WU 609/1-1. J. v. S. and A. M. R. are supported by NSF-PIF grant, by  an ARO  grant  with funding from the DARPA OLE program and by NIST.

\appendix
\section{Coulomb matrix elements}\label{App:CoulMat}

We create an orthonormal set of localized single particle orbitals $\Phi_i$ solving exactly the double dot system and we use this set to
 construct the many-body Hamiltonian for the interacting double-dot system:
\begin{multline}
  \label{HHband}
\widehat{H}=\sum_{i}E_{i}\,\, \widehat{a}^\dagger_{i} \widehat{a}_{i} +\sum_{i\ne j}T_{ij}\,\, \widehat{a}^\dagger_{i} \widehat{a}_{j}+\frac{1}{2}\sum_{ijkl} U_{ijkl}    \widehat{a}^\dagger_{i}\widehat{a}^\dagger_{j}\widehat{a}_{k} \widehat{a}_{l}.
\end{multline}
Here $E_{i}=\braket{\Phi_i|H_0|\Phi_i}$ are the single-particle energies, $T_{ij}=\braket{\Phi_i|H_0|\Phi_j}$ are the tunneling matrix elements, and $U_{ijkl}=\braket{\Phi_i \Phi_j|V_c|\Phi_k \Phi_l}$ correspond to the interaction matrix elements.
 At zero detuning, the tunneling matrix elements, $T_{ij}$, are only nonzero between states with the same $\sigma$, $\tau$, $n$ and located at different wells. For nonzero detuning, tunneling between different bands become possible.
The electrons interact through the long-range Coulomb potential, Eq.~\ref{coulomb}, and the interaction matrix elements take the form
 \begin{multline}
  \label{Uterm}
U_{nmpr}=\frac{e^2}{4 \pi^2 k_d}\int \Phi_{n}^\dagger (\zeta_1)\cdot \Phi_p(\zeta_1)\\ \Phi_{m}^\dagger (\zeta_2)\cdot \Phi_r(\zeta_2) \frac{1}{r} d\varphi_1 d\zeta_1d\varphi_2 d\zeta_2,
\end{multline}
where
\begin{equation}
  \label{CoulCyl}
\frac{1}{r}=\frac{1}{|\mathbf{r}_1-\mathbf{r}_2|}=\frac{1}{\sqrt{2R^2[1-\cos(\varphi_1-\varphi_2)]+(\zeta_1-\zeta_2)^2}}.
\end{equation}

To evaluate the interaction matrix elements, we use the following relation:
\begin{equation}
  \label{CoulCyl2}
\frac{1}{r}=\frac{1}{\pi}\int_{-\infty}^\infty e^{i q (\zeta_1-\zeta_2)} \mbox{K}_0 \big(R |2 q \sin[(\varphi_1-\varphi_2)/2]|\big)dq.
\end{equation}
If the rest of the integrand does not depend on $\varphi_1$ and $\varphi_2$, the integration over those coordinates can be done analytically,
\begin{multline}
\frac{1}{4\pi^2}\int_0^{2\pi}\int_0^{2\pi}\mbox{K}_0 \big(R |2 q \sin[(\varphi_1-\varphi_2)/2]|\big)d\varphi_1\, d\varphi_2=\\\mbox{I}_0(R|q|)\mbox{K}_0(R |q|).
\end{multline}
Also, the integration over $\zeta_1$ and $\zeta_2$ can be done analytically since $\overline{\Phi}(\zeta)$ has a simple functional dependence that can be written in terms of exponentials with real or imaginary arguments. Once those integrals are obtained, we are left with the momentum integration over $q$ in Eq.~(\ref{CoulCyl2}).
Defining
\begin{equation}
\mbox{F}_{np}(q)= \int e^{i q \zeta_1}\left\{\Phi_{n}^\dagger (\zeta_1) \Phi_p(\zeta_1)\right\} d\zeta_1,
\end{equation}
we can write the Coulomb interaction matrix as a one-dimensional numerical integration over the momentum $q$,
\begin{equation}
  \label{Uterm}
U_{nmpr}=\frac{e^2}{k_d\pi}\int \mbox{F}_{np}(q)\mbox{F}_{mr}(-q)\, \mbox{I}_0(R|q|)\mbox{K}_0(R |q|) dq.
\end{equation}
In the calculation of the Coulomb matrix elements, we neglect the dependence of the longitudinal wavevector $k_n$ on spin, valley, and magnetic field. As discussed in Sec.~\ref{SParticle}, this approximation is good for $\kappa\gg\Phi_{AB}/(\Phi_0 R),\,\Delta_{SO}/(2\hbar v)$, which is certainly true for the parameters used in this paper.

There is another effective interaction that takes into account intervalley scattering:
\begin{multline}
  \label{UKKpterm}
U^{KK'}_{nmpr}=\frac{1}{4 \pi^2}\int\Phi_{n}^\dagger (\zeta_1)\cdot \Phi_p(\zeta_1)\\ \Phi_{m}^\dagger (\zeta_2) \cdot\Phi_r(\zeta_2) V^{KK'} d\varphi_1 d\zeta_1d\varphi_2 d\zeta_2
\end{multline}
where $V^{KK'}$ is a short range interaction that can be taken proportional to $\delta(\zeta_1-\zeta_2)$. We have estimate the $U^{KK'}_{nmpr}$ terms and we have concluded that they only introduces minor corrections. For that reason, we neglect this contribution in the present work.

\section{Rate equations}\label{rate}

Here we derive the rate equations and the stationary current corresponding the the setup depicted in Fig.\ref{Blockade} a).
At zero magnetic field and large detuning the single particle ground state consists of the Kramer doublet $|L 1 K'\uparrow\rangle$ and $|L 1 K\downarrow\rangle$ localized in the left dot.
The two-particle states are characterized by their charge degree of freedom and by their spin valley symmetry.
The $|2L,0R\rangle$ and $|1L,1R\rangle$  states with antisymmetric spin-valley part are non-degenarate and are called $|M\rangle$ and $|\tilde M\rangle$ respectively. They are defined in Eq.\eqref{eqM} and \eqref{t1}.
According to Fig.\ref{table1} the $|2L,0R\rangle$ and  $|1L,1R\rangle$ states with symmetric spin-valley part are each three fold degenerate and include the states $|FF\rangle$ and $|\tilde{FF}\rangle$ respectively.
Following Eqs\eqref{t3} and \eqref{eqFF} the structure of these states is given by:
\begin{eqnarray*}
|F_-\rangle&=&\ket{FF}= |2L,0R\rangle^- \ket{K \downarrow ; K \downarrow }\\
|F_0\rangle&=& |2L,0R\rangle^-   \ket{K \downarrow ; K' \uparrow }^+\\
|F_+\rangle&=& |2L,0R\rangle^-   \ket{K' \uparrow ; K' \uparrow }\\
|\tilde{F}_-\rangle&=&\ket{\tilde{FF}}=  |1L,1R\rangle^{-}\ket{K \downarrow ; K \downarrow }\\
|\tilde{F}_0\rangle&=&  |1L,1R\rangle^{-}\ket{K \downarrow ; K' \uparrow }^+\\
|\tilde{F}_+\rangle&=& |1L,1R\rangle^{-}\ket{K' \uparrow ; K' \uparrow }\\
\end{eqnarray*}

Due to the serial setup each contact couples only to its adjacent dot.
The tunneling rates to the collector (left contact in Fig.\ref{Blockade} a)) depend on the correlated two particle states, which is accounted for by introducing the spectral weights $S_F$ and $S_M$ that determine how easy an electron tunneling between the left dot and the left reservoir can cause a transition between a two particle and one particle state on the dot.
\begin{eqnarray*}
S_F&=&\sum_m|\bra{L 1 K\downarrow} a_{LmK\downarrow}\ket{F_-}|^2\\
S_M&=&\sum_m|\bra{L 1 K\downarrow} a_{LmK'\uparrow}\ket{M}|^2\\
\end{eqnarray*}
Due to the simple form of the $|1L,1R\rangle$ wavefunctions the spectral weights corresponding to the tunneling between right contact and right dot has trivial spectral weights   $S_{\tilde{M}}=|\bra{L 1 K\downarrow} a_{R1K'\uparrow}\ket{\tilde{M}}|^2=\frac{1}{2}$ and $S_{\tilde{F}}=|\bra{L 1 K\downarrow} a_{R1K\downarrow}\ket{\tilde{F_-}}|^2=1$.

We denote the reduced density matrix for the double dot by $P$ and denote off-diagonal matrix elements with $P_{s}^{s'}:=\langle s'|P|s\rangle=(P_{s}^{s'})^*$, where $s$ and $s'$ are eigenstates of the double dot and diagonal elements with $P_{s}:=\langle s|P|s\rangle$. We neglect processes that can mix states with different spin-valley part\cite{Burkard09, Nazarov09} since hyperfine interaction is small in carbon nanotube and spin-orbit coupling is mainly a spin-valley coupling with a very weak effect on the longitudinal wave function.
Then degenerate states are equally populated and the relevant matrix elements of the reduced density matrix for the double dot system are given by $P_1=P_{|L 1 K'\uparrow\rangle}+P_{|L 1 K\downarrow\rangle}$; $P_F=P_{|F_+\rangle}+P_{|F_0\rangle}+P_{|F_-\rangle}$; $P^{\tilde{F}}_F=P^{\tilde{F}_+}_{F_+}+P^{\tilde{F}_0}_{F_0}+P^{\tilde{F}_-}_{F_-}$, etc.

The corresponding rate equations are given by:
\begin{eqnarray}
d_t P_1&=&-2 \Gamma_E P_1 +\Gamma_C S_F P_F +2 \Gamma_C S_M  P_M\\
d_t P_F &=&-\Gamma_C S_F P_F-i t_S(P_{\tilde{F}}^F-P_F^{\tilde{F}})\notag \\
d_t P_{\tilde{F}}&=&\frac{3}{2} \Gamma_E P_1+i t_S(P_{\tilde{F}}^F-P_F^{\tilde{F}})\notag\\
d_t P^{\tilde{F}}_F&=&(-\frac{1}{2}\Gamma_C S_F +i\epsilon)P^{\tilde{F}}_F-it_S(P_{\tilde{F}}-P_F)\notag\\
d_t P_M &=&-2 \Gamma_C S_M P_M-i t_{AS}(P_{\tilde{M}}^M-P_M^{\tilde{M}}) \notag\\
d_t P_{\tilde{M}}&=&\frac{1}{2} \Gamma_E P_1+i t_{AS}(P_{\tilde{M}}^M-P_M^{\tilde{M}})\notag\\
d_t P^{\tilde{M}}_M&=&-\Gamma_C S_M P^{\tilde{F}}_F-it_{AS}(P_{\tilde{M}}-P_M)\notag
\end{eqnarray}
with $P_1+P_F+P_{\tilde{F}}+P_M+P_{\tilde{M}}=1$.
Here $\Gamma_E$ denotes the coupling strength to the emitter contact (right contact in Fig.\ref{Blockade} a) and  $\Gamma_C$ denotes the coupling strength to the collector/left contact.

These rate equations are solved analytically.
The result for the stationary current and the occupation of the states are given by:
\begin{eqnarray}
I_{bl}&=&2 \frac{e \Gamma_E }{\hbar} P_1=\frac{e \Gamma_E }{\hbar} 16 S_F S_M t_S^2 t_{AS}^2 \Gamma_C/ N \label{current}\\
P_{\tilde{FF}}&=&3 S_M t_{AS}^2\Gamma_E(4t_S^2+\Gamma_C^2 S_F^2+4\epsilon^2)/N\notag\\
N&=&12 S_M t_{AS}^2\Gamma_E(2t_S^2+\epsilon^2)
+3 S_F^2 S_M t_{AS}^2 \Gamma_C^2 \Gamma_E\notag\\
& &+2 S_F t_S^2 (4 S_M t_{AS}^2 \Gamma_C+2 t_{AS}^2 \Gamma_E+S_M^2 \Gamma_C^2\Gamma_E)\notag
\end{eqnarray}
Here $t_{AS}$ denotes the interdot tunneling between states $|M\rangle$ and $|\tilde{M}\rangle$ and $t_S$ the interdot tunneling between the the states $|FF\rangle$ and $|\tilde{FF}\rangle$. Spectral weights $S_F$ and $S_M$ account for the overlap between an electron hopping onto the dot and the two particle eigenstates, and $\Gamma_C$ and $\Gamma_E$ are the coupling strength to emitter and collector.

The stationary current\eqref{current} vanishes as $1/\epsilon^2$ for large $\epsilon$. For reversed transport voltage, still a finite current can pass through the double dot which implies that, for large $\epsilon$, the double dot acts as a current rectifier.
We calculate the current for the unblocked voltage direction. Inverting the bias voltage depicted in Fig\ref{Blockade} a), the current flows by electrons hopping from the left reservoir through the double dot to the right reservoir.  Applying again the rate equation approach, we obtain:
\begin{eqnarray}
I_{op}&=&\frac{e \Gamma_R }{\hbar} P_1=\frac{2 e \Gamma_R^2 t_{AS}^2}{\hbar2 t_{AS}^2(\Gamma_R+2 S_M \Gamma_L)+S_M\Gamma_L \Gamma_R^2}  \label{current2}
\end{eqnarray}
We note that $\Gamma_L$ corresponds to $\Gamma_C$ in Eq.\eqref{current} and  $\Gamma_R$ corresponds to $\Gamma_E$.


\begin{thebibliography}{10}
\bibitem{footnote1}
 In this discussion, we have omitted the fact that, at $B=0$, the ground state of the symmetric multiplet is three fold degenerate. However, as long as spin and valley are conserved, this degeneracy does not change the physics [see Appendix~\ref{rate}].
\end{thebibliography}

\end{document}